\documentclass[10pt]{article}
\usepackage{jheppub} 
\usepackage{empheq}

\usepackage{graphicx,amsmath,amssymb,multirow,array,bm,bbm,mathrsfs}
\usepackage{relsize}%
\usepackage{verbatim}

\usepackage{epstopdf}
\usepackage{array,tikz-cd}
\newcommand*{\boxcolor}{black}
\makeatletter
\renewcommand{\boxed}[1]{\textcolor{\boxcolor}{%
\tikz[baseline={([yshift=-1ex]current bounding box.center)}] \node [rectangle, minimum width=1ex,rounded corners,draw] {\normalcolor\m@th$\displaystyle#1$};}}
 \makeatother

\definecolor{Orange}{rgb}{1,.4,0}

\newcommand{\Gstar}{{\cal G}^{(6,\,\text{star})}_T}
\newcommand{\Gbstar}{\overline{\cal G}^{(6,\,\text{star})}_{\bar{T}}}
\newcommand{\Gcomb}{{\cal G}^{(6,\,\text{comb})}_T}
\newcommand{\Vcomb}{{\cal V}^{(6,\,\text{comb})}_T}

\newcommand{\Gfour}{{\cal G}^{(4)}_T}
\newcommand{\Gfourtilde}{{\cal G}^{(4)}_{\widetilde{T}}}
\newcommand{\GfourTTtilde}{{\cal G}^{(4)}_{T/\widetilde{T}}}

\newcommand{\Gext}{\mathcal{U}_T^{(6,\text{ext.})}}
\newcommand{\Gbarext}{\overline{\mathcal{U}}_{\bar{T}}^{\rm (6, ~ext.)}}

\newcommand{\Psistar}{{\Psi}^{(6,\,\text{star})}_T}
\newcommand{\Psicomb}{{\Psi}^{(6,\,\text{comb})}_T}

\newcommand{\Psifour}{{\Psi}^{(4)}_T}

\title{On the Virasoro six-point identity block and chaos}

\author[a]{Tarek Anous}
\author[b]{and Felix M.\ Haehl}

\affiliation[\,a]{$\Delta$-Institute for Theoretical Physics \& Institute for Theoretical Physics,\\
University of Amsterdam, Science Park 904,\\
Postbus 94485, 1090 GL, Amsterdam, The Netherlands.}
\affiliation[\,b]{School of Natural Sciences, Institute for Advanced Study,\\
Einstein Drive, Princeton, NJ 08540, USA.}

\emailAdd{t.m.anous@uva.nl}
\emailAdd{haehl@ias.edu}

\abstract{We study six-point correlation functions in two dimensional conformal field theory, where the six operators are grouped in pairs with equal conformal dimension. Assuming large central charge $c$ and a sparse spectrum, the leading contribution to this correlation function is the six-point Virasoro identity block---corresponding to each distinct pair of operators fusing into the identity and its descendants. We call this the \emph{star channel}. One particular term in the star channel identity block is the stress tensor $SL(2,\mathbb{R})$ (global) block, for which we derive an explicit expression. In the holographic context, this object corresponds to a direct measure of nonlinear effects in pure gravity.
We calculate additional terms in the star channel identity block that contribute at the same order at large $c$ as the global block using the novel theory of reparametrizations, which extends the shadow operator formalism in a natural way. We investigate these blocks' relevance to quantum chaos in the form of six-point scrambling in an out-of time ordered correlator. Interestingly, the global block does not contribute to the scrambling mode of this correlator, implying that, to leading order, six-point scrambling is insensitive to the three-point graviton coupling in the bulk dual.  Finally, we compare our findings with a different OPE channel, called the comb channel, and find the same result for the chaos exponent in this decomposition.}

\begin{document} 
	
\maketitle
\flushbottom

\section{Introduction}
\label{sec:intro}

In order to understand the detailed mechanism by which gravitational dynamics is encoded in conformal field theories (CFTs), we must understand the map from basic field theory ingredients into gravity and vice versa. In a correlation function of CFT primary operators, these ``atomic'' ingredients are often the conformal blocks, which describe the exchange of a specified representation of the conformal group. Such a conformal block decomposition instructs one to pick an OPE channel and then to sum over all possible operator exchanges in that channel. The final result should be independent of which channel one chose in the beginning.

The dual gravitational computation typically looks very different. On the gravity side, in the semiclassical regime the prescription for computing correlation functions involves standard perturbative graviton exchanges in the curved geometry provided by the leading saddle point of the gravitational path integral. This calculation does not involve picking any channel. Instead it picks a leading background geometry and sums over all channels of graviton exchanges between probe operators. The standard lore is that the gravity calculation is reproduced in the CFT by assuming dominance of the identity block and maximizing its contribution over all possible identity channels \cite{Dijkgraaf:2000fq,Maloney:2016kee,Anous:2017tza}. 

In this paper we will investigate a highly nontrivial example of these rules. We study six-point functions of pairwise identical operators and assume that a suitable notion of Virasoro identity block dominates. This does not uniquely fix the decomposition since there exist topologically distinct OPE channels for six-point functions. Specifically, we mostly consider the OPE channel which we call the {\it star channel} \cite{Chen:2016dfb,Jepsen:2019svc}. As we will see, the star channel is the most direct analogue of the four-point identity block, where all external probe operators can be organized in pairs with identity monodromy. The star channel contains cubic stress tensor exchanges, and thus is genuinely sensitive to nonlinear effects in the bulk dual.

We will compare the star channel with the {\it comb channel} (see figure \ref{fig:starVScomb}), which is characterized by a different OPE channel topology \cite{Banerjee:2016qca,Alkalaev:2016rjl,Alkalaev:2018nik,Rosenhaus:2018zqn,Kusuki:2019gjs,Anous:2019yku,Fortin:2019zkm,Parikh:2019ygo,Parikh:2019dvm,Jepsen:2019svc,Alkalaev:2020kxz}. The comb channel is not a pure stress tensor block as it involves the exchange of an internal operator that is not a descendant of the identity. Thus the comb channel block is generically insensitive to the cubic graviton coupling. Of course, the final answer for the correlation function, after summing over all possible exchanges, should not depend on the particular channel we chose for our decomposition. Thus the cubic interaction of gravitons should be captured by heavy exchanges in any comb-channel decompositon.  

The correlation function we choose to study is the (maximally) out-of-time-ordered six-point correlation function (OTOC) in a thermal state. From gravitational calculations we know what to expect \cite{Larkin:1969aa,Sekino:2008he,Shenker:2013pqa,Shenker:2013yza,Leichenauer:2014nxa,Maldacena:2015waa,Kitaev:2015aa,Haehl:2017pak}: the OTOC should display black hole-like features and in particular it should have an exponentially growing contribution, which signals scrambling at the horizon. We indeed find such a contribution in both star and comb identity blocks. Interestingly, the term responsible for this growing mode in the star channel is \emph{not} the global $T$ block. 

This paper touches upon a second topic of recent interest. It was pointed out that a theory of reparametrization modes in two-dimensional CFTs is useful for understanding identity blocks and quantum chaos \cite{Turiaci:2016cvo,Haehl:2018izb,Cotler:2018zff}. One way to think about the theory of reparametrizations is in terms of the geometric action describing the quantization of coadjoint orbits of two copies of Diff$(S^1)/SL(2,\mathbb{R})$ \cite{Witten:1987ty,Alekseev:1988ce,Rai:1989js,Verlinde:1989ua}. In \cite{Haehl:2019eae} it was observed that at the linearized level this theory is in direct correspondence with the shadow operator formalism \cite{Dolan:2011dv,SimmonsDuffin:2012uy} applied to compute contributions of stress tensor exchanges to global conformal blocks. While we will not say much about the reparametrization mode perspective, we will make extensive use of the technical simplifications it brings about when phrased in terms of the shadow operator approach to global blocks. A novel development in the present paper is the computation of the reparametrization mode three-point function, see \eqref{eq:eps3phys}. It is the basic ingredient in our derivation of the star channel identity block alluded to above. This calculation illustrates clearly some of the advantages of the reparametrization mode formalism for computing stress tensor blocks.

This paper is organized as follows. In \S\ref{sec:blocks} and \S\ref{subsec:blockdefs} we discuss the different OPE channel topologies and define the global six-point star-channel $T$ block. We derive this block using the shadow operator formalism and reparametrization modes in \S\ref{sec:derivation}. In \S\ref{sec:virasoro} we generalize the formalism in order to discuss the Virasoro block in certain kinematic regimes. In \S\ref{sec:chaos} we analytically continue the star-channel identity block to the second sheet such as to obtain a particular out-of-time-ordered six-point function. We show that the global block is not sufficient to capture interesting out-of-time-order dynamics, but the first nontrivial piece in the Virasoro block in fact dominates and allows us to identify the relevant six-point scrambling time.  We compare our results with the similar looking block derived in the comb channel, see \S\ref{sec:comb}. We end with a discussion in \S\ref{sec:conclusions} and defer some technical details to appendices.\\

{\it Note:} While this paper was nearing completion, ref.\ \cite{Fortin:2020yjz} appeared, which has partial overlap with some of our discussion of the global star (which they refer to as \emph{snowflake}) and comb channel blocks. 

\section{The star-channel global \texorpdfstring{$T$}{T} block}
\label{Sec:globalblocks}


\subsection{Review of the conformal block expansion}\label{sec:blocks}

In this paper we consider 2d CFT correlation functions of six operators, grouped in pairs: 
\begin{multline}\label{eq:gencorr}
	G(x_1,\bar{x}_1,x_2,\bar{x}_2,y_1,\bar{y}_1,y_2,\bar{y}_2,w_1,\bar{w}_1,w_2,\bar{w}_2)=\\\langle X(x_1,\bar{x}_1)X(x_2,\bar{x}_2) Y(y_1,\bar{y}_1)Y(y_2,\bar{y}_2)W(w_1,\bar{w}_1)W(w_2,\bar{w}_2)\rangle~.
\end{multline}
Generally, the operators have dimensions $\Delta_X=h_X+\bar{h}_X$, $\Delta_Y=h_Y+\bar{h}_Y$ and  $\Delta_W=h_W+\bar{h}_W$ and spin $s_X=h_X-\bar{h}_X$, $s_Y=h_Y-\bar{h}_Y$ and  $s_W=h_W-\bar{h}_W$~. However, for the remainder of the paper we will focus on scalar operators with $s_{X/Y/Z}=0$~. Since we work in two dimensions, this is not a very restrictive choice and generalizations are straightforward.

For scalar operators we can express the above correlation function as follows 
\begin{equation}
 	G=\frac{1}{|x_{12}|^{2\Delta_X}|y_{12}|^{2\Delta_Y}|w_{12}|^{2\Delta_W}}F(z,\bar{z},u,\bar{u},v,\bar{v})
 \end{equation} 
 where $z_{ij}\equiv z_i-z_j$ and the function $F$ only depends on the following conformally invariant cross-ratios
 \begin{equation}
\label{eq:crossDef}
   z = \frac{(x_1-y_1)(y_2-x_2)}{(x_1-y_2)(y_1-x_2)} \,, \qquad u = \frac{(x_1-y_1)(y_2-w_1)}{(x_1-y_2)(y_1-w_1)} \,, \qquad v =  \frac{(x_1-y_1)(y_2-w_2)}{(x_1-y_2)(y_1-w_2)}\,,
\end{equation}
and their anti-holomorphic counterparts. Two additional simplifications happen in two-dimensional conformal field theory. By associativity of the operator algebra, the function $F$ can be decomposed as an infinte sum over intermediate exchanges of operators. The functions labeling individual exchanges are known as \emph{conformal blocks} and by holomorphy of 2d CFT, they decompose into a product of a holomorphic function and an anti-holomorphic function. In summary, we may write: 
\begin{equation}\label{eq:decomp}
	F(z,\bar{z},u,\bar u, w,\bar w)=\sum_{i,j,\dots}c_{i,j,\dots}\bar{c}_{i,j,\dots}\mathcal{V}_{i,j,\dots}(z,u,v)\bar{\mathcal{V}}_{i,j,\dots}(\bar{z},\bar{u},\bar{v})
\end{equation}
where the coefficients $c_{i,j,\dots}$ denote products of three-point coefficients and are theory dependent. The functions $\mathcal{V}(z,u,v)$ are the individual blocks, which only contain kinematic data. 

In 2d CFT these functions $\mathcal{V}$ label the possible exchanges of 
\emph{Virasoro primary} operators and all their descendents, thus containing the information of an entire Virasoro representation. However, without specifying the microscopic data of our CFT, it is difficult to proceed beyond \eqref{eq:decomp}. In theories with gravity duals, on the other hand, we may compute certain universal contributions to the correlator \eqref{eq:gencorr}, since we have come to expect the Virasoro block associated with the identity operator to dominate in certain kinematics \cite{Yin:2007gv,Headrick:2010zt,Hartman:2013mia} as these have the universal features necessary to reproduce bulk AdS$_3$ physics, including multi-graviton exchanges \cite{Fitzpatrick:2014vua}. This has recently been used in a wide range of applications related to 3d black hole spacetimes \cite{Anous:2016kss,Fitzpatrick:2016ive,Anous:2017tza}. 

Even computing the individual Virasoro blocks is a difficult task, although recursion relations exist for computing them order by order in a small cross-ratio expansion \cite{Zamolodchikov:1985ie,zamolodchikov1987conformal}. At large-$c$, the blocks exponentiate and can be obtained using the monodromy method \cite{Zamolodchikov:1987ae}. 

\begin{figure}
	\centering     
	\includegraphics[width=.7\textwidth]{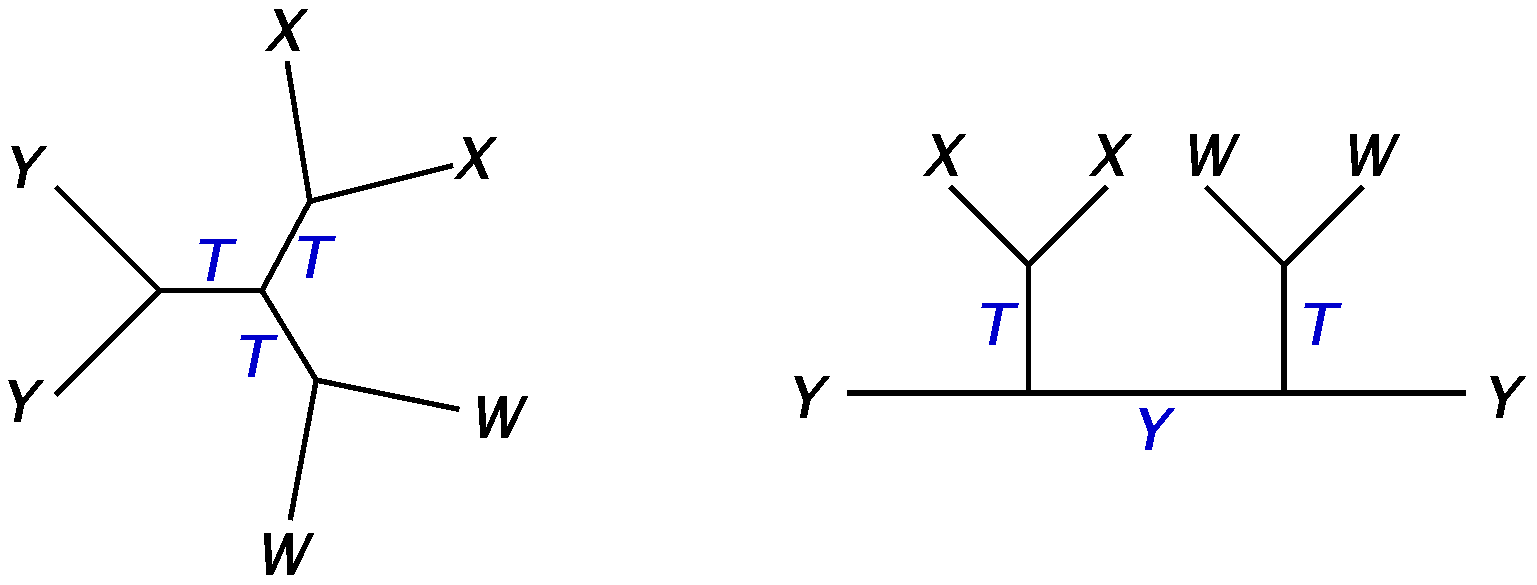}
	\caption{Illustration of the two different OPE channel topologies relevant for universal contributions to the six-point block of pairs of operators $X,Y,W$. The left figure shows the {\it star channel} identity block, which involves purely stress tensor exchanges. For comparison, on the right hand side we also show the {\it comb channel}, which involves an intermediate exchange that is identical to one of the external probe operators.}
	\label{fig:starVScomb}
\end{figure}

For four- and five-point functions, there is a unique choice of graph topology for the conformal block expansion.\footnote{ Of course there exist different channels for four-point functions, but we stress that they all have the same topology.} This topology is known as the \emph{comb}. However at six points, we are faced with a choice between two types of graph topologies, both denoted in figure \ref{fig:starVScomb}. The new type of graph is called the \emph{star} (or snowflake \cite{Fortin:2020yjz}) and is shown in the left hand side of the figure, while the comb is drawn in the right side of the figure.

While the specific comb channel shown in the figure is still universal, it is clear that the most direct analogue of the four-point Virasoro identity block---where pairs of external operators with the same dimension are taken to fuse into the identity---is given by the star channel OPE. As we will see, the leading contribution to the star channel Virasoro identity block contains the star channel global block plus additional terms. As previously noticed \cite{Fitzpatrick:2016thx}, in the case of four-point functions, knowledge of these global $T$ blocks is enough for extracting the chaos exponent in theories with a gravity dual. We will show (see \S\ref{sec:chaos}) that the star-channel global $T$ block is not entirely sufficient for extracting the chaos exponent at six points.  

In this note we will always refer to  Virasoro blocks as $\mathcal{V}$ and denote global blocks by $\mathcal{G}$. We shall carefully distinguish the global from the Virasoro block in what follows. It is very instructive to first develop methods for computing the global block and then generalize them for the Virasoro case.

\paragraph{Characterization of the global star $T$ block:}
The global block $\Gstar$ will make an appearance in what follows, so we characterize it here to make its identification more obvious. Since $\Gstar$ is fixed by global conformal symmetry, it satisfies a set of differential equations known as the conformal Casimir equations (see e.g. \cite{Dolan:2003hv}). 
For the sake of pedagogy, we present here the Casimir equations satisfied by the star channel block in the left of figure \ref{fig:starVScomb}:
\begin{align}
&\left[-(b-c)^2\partial_b\partial_c-h_T(h_T-1)\right]\Gstar(z,u,v)=0\label{eq:cas1}\\
&\left[(1-z)^2\left\lbrace \partial_z z\partial_z+\left(u\,\partial_u+v\,\partial_v\right)\partial_z\right\rbrace-h_T(h_T-1)\right]\Gstar(z,u,v)=0\label{eq:cas2}\\
&\left[-(u-v)^2\partial_u\partial_v-h_T(h_T-1)\right]\Gstar(z,u,v)=0\label{eq:cas3}
\end{align}
with $h_T=2$ (for stress tensor exchanges) and $b\equiv\frac{u}{v}\frac{1-v}{1-u}$ and $c\equiv\frac{1-v}{1-u}$. We will review how to derive these equations in section \ref{subsec:blockdefs}.

\subsection{The shadow operator formalism and definition of the star channel global block}
\label{subsec:blockdefs}

Consider the conformal six-point function $G$ as denoted in \eqref{eq:gencorr}. To decompose it into conformal blocks, we must insert three complete sets of states (or resolutions of the identity) and obtain sums over products of three-point functions. For the global blocks, we shall employ the shadow operator formalism \cite{Dolan:2011dv,SimmonsDuffin:2012uy}, where projection onto the conformal family of a primary ${\cal O}$ and its global descendants is implemented by the conformally invariant projector 
\begin{equation}
\label{eq:projDef}
    |{\cal O}| 
    \equiv \frac{(2h-1)(1-2\bar{h})}{C_\mathcal{O}\pi^2}  \int d^2z d^2z' \; |{\cal O}(z) \rangle \,\frac{1}{(z-z')^{2-2h} (\bar{z}-\bar{z}')^{2-2\bar{h}}}\, \langle  {\cal O}(z') |
\end{equation}
where $(h,\bar{h})$ are the dimensions of ${\cal O}$ and $C_\mathcal{O}$ is the normalization of $\langle \mathcal{O}\mathcal{O}\rangle$. We will generally work with $C_\mathcal{O}=1$ except when $\mathcal{O}=T$, in which case $C_T=c/2$.\footnote{ The normalization in \eqref{eq:projDef} is chosen such that $|{\cal O}|^2 = |{\cal O}|$.} 
The projector can also be written in terms of the formal shadow operator\footnote{ We leave spin indices implicit. For instance, operators in \eqref{eq:shadowDef} should be understood as ${\cal O} \equiv {\cal O}_{z \cdots z}$ and $\widetilde{\cal O} \equiv \widetilde{\cal O}^{z \cdots z}$.}
\begin{equation}
\label{eq:shadowDef}
\widetilde{\cal O}(z) \equiv \frac{(-2)^{h-\bar{h}}\,\Gamma(2-2\bar{h})}{\pi\Gamma(2h-1)} \int d^2z \; \frac{1}{(z-z')^{2-2h} (\bar{z}-\bar{z}')^{2-2\bar{h}}}\; {\cal O}(z') \,,
\end{equation}
which has dimensions $\big(\widetilde{h},\bar{\widetilde{h}}\big) = (1-h,1-\bar{h})$.  
Exchanges of the stress tensor $T \equiv T_{zz}$ are obtained by using the stress tensor projector
\begin{equation}
\label{eq:TprojDef}
  |T| = \frac{6}{\pi^2c} \int d^2z d^2z' \; |T(z)\rangle \, \frac{(z-z')^2}{(\bar{z}-\bar{z}')^2} \, \langle T(z') | \equiv \frac{3}{\pi c} \int d^2z \; |T(z)\rangle \langle \widetilde{T}(z) |  \,,
\end{equation}
where in the second equality we have used the definition of the shadow stress tensor 
\begin{equation}\label{eq:TtildeDef}
	\widetilde{T}(z)\equiv \widetilde{T}^{zz} = \frac{2}{\pi}\int d^2z' \; \frac{(z-z')^2}{(\bar{z}-\bar{z}')^2} \, T(z')~.
\end{equation}
Inserting these projectors into a correlation function will not produce an isolated global block. It instead produces a conformal partial wave (CPW), which we shall hereafter denote by $\Psi$. The CPW $\Psi$ is a linear combination of the block associated with ${\cal O}$ exchange plus the shadow block associated with the exchange of a shadow representation $\widetilde{\cal O}$. Since both of these share the same Casimir eigenvalues, they are instead distinguished by their short distance behavior. To obtain a global block from a CPW, we must further project out the shadow representation. This is done via an additonal \emph{monodromy projection} and below, we will implement the monodromy projection in an efficient way.

The star channel CPW is defined by fusing each pair of operators into a stress tensor, i.e., 
\begin{equation}
   X(x_1)X(x_2) \; \longrightarrow \; \frac{3}{\pi c} \int d^2z \; \langle X(x_1)X(x_2){T}(z) \rangle \; \widetilde{T}(z)
\end{equation}
and similarly for $Y$ and $W$ operators. Using this OPE structure symmetrically in the six-point function, we obtain the star channel CPW:
\begin{equation}\label{eq:f6def}
   \Psistar \equiv  \left(\frac{3}{\pi c}\right)^3 \iiint d^2z_a\, d^2z_b\, d^2z_c\;  \frac{\big{\langle} X_1 X_2 {T}_a \big{\rangle}}{ \langle X_1X_2 \rangle} \frac{\big{\langle}  Y_1 Y_2 T_b\big{\rangle}}{ \langle Y_1Y_2 \rangle} \frac{  \big{\langle}  W_1 W_2 T_c \big{\rangle} }{ \langle W_1W_2\rangle} \;\big{\langle} \widetilde{T}_a \widetilde{T}_b \widetilde{T}_c \big{\rangle}~.
\end{equation}
We illustrate the OPE structure in the left panel of figure \ref{fig:starVScomb}. The star channel has been discussed in \cite{Chen:2016dfb,Jepsen:2019svc}.
Our first goal is to give an explicit expression for the block $\Gstar$ describing external operators fusing pairwise into  $T$.

As mentioned earlier, the CPW contains the block of interest as well as a shadow block. The six-point global blocks are obtained from the CPWs by performing a monodromy projection, which we denote abstractly as follows:
\begin{equation}
   \Gstar = \Psistar \big{|}_{\rm phys.}  \,,
\end{equation}
and will provide the details of this procedure in section \ref{sec:derivation}, however, let us give a rough intuition of the procedure now. In the next section, following \cite{Haehl:2019eae}, we will give Feynman rules for computing CPWs with stress tensor exchanges. These Feynman rules are built out of particular propagators, and similarly to the computation of correlation functions in any quantum field theory, there are various propagators one can choose (e.g. Feynman/retarded/advanced) which give rise to different observables, in practice. The monodromy projection we discuss is simply a choice of propagator which results in a block rather than a CPW.

\paragraph{Casimir equations:} As alluded to above, the global six-point blocks can be characterized as solutions to certain Casimir equations, which generally have the form :
\begin{equation}
	\left[\mathcal{C}(1,\dots,k)-h_{\rm ex}(h_{\rm ex}-1)\right]\mathcal{G}=0
\end{equation}
when $k$ external operators fuse into an internal operator of dimension $h_{\rm ex}$ and $\mathcal{C}(1,\dots,k)$ is a particular quadratic Casimir of the conformal group.
In two dimensions these  quadratic Casimirs can be constructed as follows. Define the action of (holomorphic) generators on some operator ${\cal O}(z_i)$ with holomorphic weight $h_i$ as:
\begin{equation}
\qquad L_{-1}^{(i)} = \partial_i \,, \qquad  L_0^{(i)} = z_i \partial_i + h_i \,, \qquad L_1^{(i)} = z_i^2 \partial_i +2h_i z_i  \,.
\end{equation}
The quadratic Casimir acting on $k$ external operators is 
\begin{equation}
  {\cal C}(1,\ldots,k) = \Big(\mathsmaller{\sum}_i \, L_0^{(i)} \Big)\left( \mathsmaller{\sum}_i \, L_0^{(i)} \right) - \frac{1}{2} \left( \mathsmaller{\sum}_i \, L_1^{(i)} \right)\left( \mathsmaller{\sum}_i \,  L_{-1}^{(i)} \right) -  \frac{1}{2} \left( \mathsmaller{\sum}_i\,  L_{-1}^{(i)} \right)\left( \mathsmaller{\sum}_i \, L_{1}^{(i)} \right) 
\end{equation}
where sums run over $i=1,\ldots,k$. 
In particular, we have:
\begin{align}
\label{eq:casimirsDef}
   {\cal C}(z_1) &= h_1(h_1-1)  \,,\nonumber\\
   {\cal C}(z_1,z_2) &= (h_1+h_2)(h_1+h_2-1)  - z_{12}^2 \, \partial_1 \partial_2+2 \,z_{12} \left( h_2\, \partial_1 - h_1\, \partial_2 \right)  \,,\\
   {\cal C}(z_1,z_2,z_3) &= (h_1+h_2+h_3)(h_1+h_2+h_3-1)- z_{12}^2\, \partial_1 \partial_2 - z_{13}^2\, \partial_1 \partial_3 - z_{23}^2\, \partial_2 \partial_3 \nonumber\\
   &\quad + 2z_{12}(h_2 \,\partial_1 - h_1 \,\partial_2) + 2z_{23}(h_3 \,\partial_2 - h_2 \,\partial_3) + 2z_{31}(h_1 \,\partial_3 - h_3 \,\partial_1)  \,.\nonumber
\end{align}

Since the six-point blocks depend on three cross-ratios, there are also three independent Casimir equations, which should be satisfied simultaneously. In the star channel, all pairs of operators $X_{1,2}$, $Y_{1,2}$, and $W_{1,2}$ fuse into stress tensor exchanges. Therefore, the CPW and the global part of the block satisfy:
\begin{align}
 {\cal C}(x_1,x_2) \left[ \langle XX \rangle \langle YY \rangle \langle WW\rangle  \, \Psistar \right] &= 2 \, \langle XX \rangle \langle YY \rangle \langle WW\rangle \, \Psistar  \label{eq:6ptCasimir1}\\
   {\cal C}(y_1,y_2) \left[ \langle XX \rangle \langle YY \rangle \langle WW\rangle  \, \Psistar  \right] &= 2 \, \langle XX \rangle \langle YY \rangle \langle WW\rangle \, \Psistar \label{eq:6ptCasimi2}\\
   {\cal C}(w_1,w_2) \left[ \langle XX \rangle \langle YY \rangle \langle WW\rangle  \, \Psistar \right] &= 2 \, \langle XX \rangle \langle YY \rangle \langle WW\rangle \, \Psistar
   \label{eq:6ptCasimi3}
\end{align}
where $h_T(h_T-1)=2$ is the eigenvalue associated with fusion of holomorphic external operators into stress tensors and e.g. $\langle XX \rangle$ is shorthand for $(x_1-x_2)^{-2h_X}$~.

\subsection{Derivation of the star-channel six-point \texorpdfstring{$T$}{T} block}
\label{sec:derivation}

In this section we show how to derive the result for the star channel block (see \eqref{eq:G6res}). We phrase the computation in terms of the shadow operator formalism applied to stress tensor exchanges. Note, however, that due to the close relation between stress tensor shadows and reparametrization modes in CFTs \cite{Haehl:2019eae}, the computations below find a natural home in the context of the theory of reparametrizations developed in \cite{Haehl:2018izb,Cotler:2018zff,Haehl:2019eae,Jensen:2019cmr}.

\subsubsection{Four-point blocks from the shadow operator formalism}
\label{sec:...}

To set the stage, we begin with a review of the global stress tensor four-point block both from the perspective of the shadow operator formalism, and in terms of reparametrization modes. We will briefly comment on the four-point Virasoro identity block. 

We can define the global stress tensor four-point block using the shadow operator formalism. We begin by repeating expression \eqref{eq:TtildeDef} for the shadow of the holomorphic stress tensor \cite{Dolan:2011dv,SimmonsDuffin:2012uy}:
\begin{equation}
\label{eq:TtildeDef2}
   \widetilde{T}(z) =\frac{2}{\pi}\int d^2 z' \; \frac{(z-z')^2}{(\bar{z}-\bar{z}')^2}  \; T(z') \,.
\end{equation}
A central purpose of this formal definition is that it allows us to define the projectors onto stress tensor blocks, \eqref{eq:TprojDef}.
We can use these to project the four-point function onto the exchange of a stress tensor and its global descendants.  This defines the following four-point CPW:
\begin{equation}
\label{eq:4ptCPW}
\begin{split}
  & \Psifour(y_1,y_2,w_1,w_2)  \equiv   \frac{\big{\langle} Y_1 Y_2 | {T} | W_1W_2 \big{\rangle}}{\langle Y_1Y_2 \rangle \langle W_1W_2\rangle} 
\end{split}
\end{equation}
where we assume for simplicity that $Y$ and $W$ are purely holomorphic operators. According to \eqref{eq:TprojDef} and \eqref{eq:TtildeDef2}, the computation of this object involves two conformal integrals.
One can perform these explicitly \cite{Dolan:2011dv} and obtains the following result:
\begin{equation}
\label{eq:f4res}
\begin{split}
   \Psifour(z,\bar{z}) & = \frac{2h_Yh_W}{ c}  \, \left[ z^2 \; {}_2F_1(2,2,4,z) + 12 \, \frac{\bar{z}}{z} \; {}_2F_1(-1,-1,-2,z) \; {}_2F_1(1,1,2,\bar{z}) \right]
   \end{split}
\end{equation}
where $z \equiv \frac{(y_1-y_2)(w_1-w_2)}{(y_1-w_1)(y_2-w_2)}$.
This CPW is the sum of the well-known global four-point block $\Gfour(z)$ and the  shadow block $\Gfourtilde(z,\bar{z})$. Both solve the four-point Casimir equation,
\begin{equation}
\label{eq:Casimir4pt}
    {\cal C}(y_1,y_2) \left[ \langle Y_1Y_2 \rangle \langle W_1W_2\rangle \, \GfourTTtilde  \right] = 2 \, \langle Y_1Y_2 \rangle \langle W_1W_2\rangle \, \GfourTTtilde  \,,
\end{equation}
but only $\Gfour$ has the desired short distance behavior corresponding to stress tensor exchange. The monodromy projection amounts to dropping the shadow block from the CPW \cite{SimmonsDuffin:2012uy,Fitzpatrick:2011hu}. This leaves us with the global stress tensor block:
\begin{equation}
    \Gfour(z) \equiv \Psifour(z,\bar{z}) \Big{|}_{\rm phys.} =  \frac{2h_Yh_W}{ c}  \;  z^2 \; {}_2F_1(2,2,4,z)  \,.
\end{equation}

\paragraph{Formulation in terms of reparametrization modes:}
Let us now recall how the above calculation can be phrased in terms of the exchange of a nonlocal operator of negative dimension (closely related to a holomorphic reparametrization mode). We will refer to this as the reparametrization mode method. The following discussion is mostly a review of \cite{Haehl:2019eae}, building partly on previous work \cite{Haehl:2018izb,Cotler:2018zff}. At the level of global blocks, the reparametrization mode method is equivalent to the shadow operator formalism, but simpler to perform in practice for at least three reasons: 
\begin{enumerate}
\item The reparametrization mode calculation is local in the sense that it eliminates the need to perform any conformal integrals.
\item The monodromy projection can be performed at the level of the reparametrization mode propagator. We can then compute the physical conformal block directly without having to extract it from the conformal partial wave.
\item The fact that the block satisfies Casimir equations will be manifest in the calculation.
\end{enumerate}
While these simplifications are minor in case of the global four-point block with $T$ exchanged, they will be crucial when we turn to the six-point case.\footnote{ Some simplifications due to the reparametrization mode formalism also become apparent for four-point blocks in higher dimensions \cite{Haehl:2019eae}.}

In order to write the above calculation in terms of reparametrization modes, let us write the shadow of the stress tensor as a derivative:\footnote{ The normalization is arbitrary, but will ensure consistency with the formalism used in section \ref{sec:virasoro}.}
\begin{equation}
\label{eq:TtEps}
  \widetilde{T} = \frac{c}{3} \, \bar{\partial}\epsilon  \,.
\end{equation}
We will often refer to $\epsilon$ as the reparametrization mode. This is based on the observation that its correlation functions can be obtained from an effective action of holomorphic reparametrizations. We refer the reader to \cite{Haehl:2018izb,Cotler:2018zff} for a more in depth explanation. At present, we simply treat $\epsilon$ as an operator with dimensions $(h,\bar{h}) = (-1,0)$, which can be defined nonlocally in terms of the stress tensor through \eqref{eq:TtEps} and \eqref{eq:TtildeDef2}.

We can now rewrite the four-point stress tensor CPW \eqref{eq:4ptCPW} as follows:
\begin{equation}
\label{eq:f4Calc}
\begin{split}
   \Psi_T^{(4)} 
    &= \frac{1}{\langle Y_1Y_2\rangle\langle W_1W_2\rangle} \left( \frac{3}{\pi c}\right)^2 \int d^2z_a\, d^2z_b\; \big{\langle} Y_1Y_2 {T}_a \big{\rangle} \big{\langle} \widetilde{T}_a \widetilde{T}_b \big{\rangle} \big{\langle} T_b W_1W_2 \big{\rangle} \\
    &= \frac{1}{\langle Y_1Y_2\rangle\langle W_1W_2\rangle} \frac{1}{\pi^2} \int d^2z_a\, d^2z_b \; \bar{\partial}_a\big{\langle} T_aY_1Y_2  \big{\rangle} \big{\langle} \epsilon_a \epsilon_b \big{\rangle} \, \bar{\partial}_b\big{\langle} T_b W_1W_2 \big{\rangle}
   \end{split}
\end{equation}
Note that the two conformal integrals are now trivial to do. We simply use the conformal Ward identity 
\begin{equation}
  \bar{\partial}_a \big{\langle} T_aY_1Y_2  \big{\rangle}  = - \pi  \,  \sum_{i={1,2}} \left( h_Y\, \partial_{z_a}  \delta^{(2)}(z_a-y_i)  -  \delta^{(2)}(z_a-y_i) \partial_{y_i} \right)\, \langle Y_1Y_2 \rangle
\end{equation} 
to eliminate the integrals. We can write this as:
\begin{equation}
\label{eq:B1B1}
\Psi_T^{(4)} = \big{\langle}  {\cal B}^{(1)}_{\epsilon,h_Y}(y_1,y_2)\,  {\cal B}^{(1)}_{\epsilon,h_W}(w_1,w_2) \big{\rangle} \; \equiv
 \;\;
\begin{gathered}
\includegraphics[width=.07\textwidth]{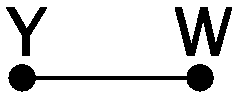}
\end{gathered}
\end{equation}
where we have introduced diagrammatic notation to indicate a single $\epsilon$-exchange (similar to diagrams in \cite{Qi:2019gny}) between two bilocal operators $\mathcal{B}^{(1)}_{\epsilon,h}$: each bilocal can be thought of as a local operator (or {\it OPE block}) in kinematic space \cite{Czech:2016xec,deBoer:2016pqk} and is therefore indicated by a single dot. We note here the superscript $(1)$, but defer its meaning to section \ref{sec:virasoro}. The $\epsilon$-exchange is depicted by a line in the diagram, and the bilocal is defined as
\begin{equation}
\label{eq:B1def}
   {\cal B}_{\epsilon,h}^{(1)} (z_1,z_2) \equiv  
h \left( \partial \epsilon(z_1) +\partial \epsilon(z_2)  -2\, \frac{\epsilon(z_1) - \epsilon(z_2)}{z_1-z_2}\right) \,.
\end{equation}
This diagram is more of a mnemonic than an actual Feynman diagram, as the bilocals themselves also depend on $\epsilon$.

With this formalism introduced, the four-point CPW with $T$-exchanges can be expressed as a two-point function of these bilocal reparametrization mode insertions. All we need in order to evaluate this object is the Euclidean two-point function of the reparametrization mode $\epsilon$. This can be reverse-engineered from the two-point function of the stress tensor shadow: 
\begin{equation}
\label{eq:constr1}
\langle \widetilde{T}_1 \widetilde{T}_2 \rangle=\frac{2c}{3} \frac{z_{12}^2}{\bar{z}_{12}^2}\equiv  \left(\frac{c}{3}\right)^2 \langle\bar{\partial}\epsilon_1\, \bar{\partial} \epsilon_2 \rangle
\end{equation}
where $\epsilon_i = \epsilon(z_i,\bar{z}_i)$ etc. 
Note that due to the relation\footnote{ Recall that $\bar{\partial} \frac{1}{z} = \pi \, \delta^{(2)}(z)$.}
\begin{equation}
\label{eq:TEps}
T(z) = \widetilde{\widetilde{T}}(z) = \frac{3}{\pi c} \int d^2 z' \; \langle T(z) T(z') \rangle \, \widetilde{T}(z')
=- \frac{1}{\pi} \int d^2 z' \; \bar{\partial}' \langle T(z)  T(z') \rangle \, \epsilon(z') =-  \frac{c}{12} \, \partial^3 \epsilon(z)\,,
\end{equation}
we obtain further constraints on derivatives of the $\langle\epsilon\epsilon\rangle$ two-point function:
\begin{equation}
\label{eq:constr2}
\langle{\partial}^3\epsilon_1\, {\partial}^3 \epsilon_2 \rangle = \frac{144}{c^2}\, \langle {T}_1 {T}_2 \rangle = \frac{72}{c} \frac{1}{{z}_{12}^4}   \,,\qquad
\langle \bar{\partial}\epsilon_1\, {\partial}^3 \epsilon_2 \rangle = - \frac{36}{c^2} \, \langle \widetilde{T}_1 T_2 \rangle = -\frac{12\pi}{c} \, \delta^{(2)}(z_{12})\,.
\end{equation}
The constraints \eqref{eq:constr1} and \eqref{eq:constr2} can easily be solved by integration. This gives:
\begin{equation}
\label{eq:epsprop}
 \langle \epsilon(z_1,\bar{z}_1)\epsilon(z_2,\bar{z}_2) \rangle 
 = \frac{6}{c} \, (z_1-z_2)^2 \, \log|z_1 - z_2|^2 
\end{equation}
We could add further ``integration constants'' to the result \eqref{eq:epsprop} (such as a purely quadratic term $\propto z_{12}^2$), but these would be superfluous in the sense that they do not contribute to any physical correlation functions.
One can immediately verify that the evaluation of \eqref{eq:B1B1} using this propagator reproduces the global CPW, \eqref{eq:f4res}.

The reparametrization mode formulation gives us another advantage: we can perform the monodromy projection at the level of the $\epsilon$-propagator such that the exchange \eqref{eq:B1B1} yields the physical conformal block directly.
To this end, note that it is natural to split this propagator into the sum of a physical and a shadow part \cite{Haehl:2019eae}:
\begin{equation}
\label{eq:epspropSplit}
\begin{split}
\langle \epsilon_1 \epsilon_2  \rangle = \langle \epsilon_1 \epsilon_2  \rangle_{\rm phys.} + \langle \epsilon_1 \epsilon_2  \rangle_{\rm shad.} \;\; & \\
\text{where:}\quad 
 \langle \epsilon(z_1,\bar{z}_1)\epsilon(z_2,\bar{z}_2) \rangle_{\rm phys.} 
 &= \frac{6}{c} \, (z_1-z_2)^2 \, \log(z_1 - z_2)  \\
  \langle \epsilon(z_1,\bar{z}_1)\epsilon(z_2,\bar{z}_2) \rangle_{\rm shad.} 
 &= \frac{6}{c} \, (z_1-z_2)^2 \, \log(\bar{z}_1 - \bar{z}_2) 
\end{split}
\end{equation}
Using the physical propagator in \eqref{eq:B1B1} produces the physical stress tensor conformal block. On the other hand, the shadow block is computed by using the shadow propagator in \eqref{eq:B1B1} (this is explained in detail in \cite{Haehl:2019eae}).

Finally, let us discuss the four-point Casimir equation \eqref{eq:Casimir4pt} in the reparametrization mode language. Note the following identity: 
\begin{equation}
\label{eq:Casimir4ptB1}
    {\cal C}(y_1,y_2) \left[ \langle Y_1Y_2 \rangle \;  {\cal B}^{(1)}_{\epsilon, h_Y}(y_1,y_2) \right] = 2 \, \langle Y_1Y_2 \rangle \; {\cal B}^{(1)}_{\epsilon, h_Y}(y_1,y_2)\,.
\end{equation}
That is, the bilocal ${\cal B}^{(1)}_{\epsilon,h}$ is itself an eigenfunction of the Casimir with eigenvalue corresponding to stress tensor exchange.\footnote{ In the kinematic space picture, the Casimir acts as a wave operator and \eqref{eq:Casimir4ptB1} is interpreted as a wave equation for the OPE block \cite{Czech:2016xec,deBoer:2016pqk}.} It is therefore manifest that the CPW as computed in the reparametrization mode formalism, \eqref{eq:B1B1}, solves the defining Casimir equation. This remains true for any choice of reparametrization mode propagator (the full propagator \eqref{eq:epsprop}, its physical piece, or its shadow piece).

\subsubsection{Star channel six-point block from reparametrization modes}

Having reviewed the computation of the global four-point block, we now turn to the star channel six-point block where the advantages of the formalism presented become apparent.

 We are confronted with two problems when evaluating the CPW $\Psistar$ as defined in \eqref{eq:f6def}: first, just performing these conformal integrals is very tedious. Second, once the integrals have been done, one still needs to identify the physical block and project out the shadow contributions. Note that now there will be $5$ shadow blocks in addition to the one physical block. 

We will circumvent these issues by using the simplifications in stress tensor exchanges due to Ward identities. Using the same arguments as for the four-point block, it is immediately clear that the evaluation of \eqref{eq:f6def} reduces to the following three-point function of bilocal operators:
\begin{equation}
\Psistar = \big{\langle}  {\cal B}^{(1)}_{\epsilon,h_X}(x_1,x_2)\,  {\cal B}^{(1)}_{\epsilon,h_Y}(y_1,y_2)\,{\cal B}^{(1)}_{\epsilon,h_W}(w_1,w_2) \big{\rangle}    \; = \;\; 
  \begin{gathered}\vspace{-.2cm}
\includegraphics[width=.09\textwidth]{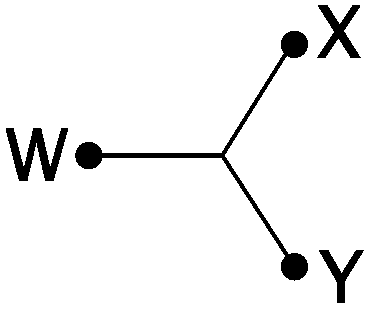}
\end{gathered}
\end{equation}
This calculation becomes straightforward once we figure out the three-point function $\langle \epsilon \epsilon \epsilon \rangle$. Note that this will a priori again give a CPW rather than the physical block. The latter is obtained by means of a monodromy projection:
\begin{equation}
\label{eq:G6def}
\begin{split}
   \Gstar (z,u,v) &\equiv \Psistar (z,u,v)\Big{|}_{\rm phys.}= \big{\langle}  {\cal B}^{(1)}_{\epsilon,h_X}(x_1,x_2)\,  {\cal B}^{(1)}_{\epsilon,h_Y}(y_1,y_2)\,{\cal B}^{(1)}_{\epsilon,h_W}(w_1,w_2) \big{\rangle}_{\rm phys.}
\end{split}
\end{equation}
We will momentarily define the right hand side by deriving $\langle \epsilon \epsilon \epsilon \rangle_{\rm phys.}$.

To obtain the  three-point function $\langle \epsilon \epsilon \epsilon \rangle$, we can follow the same strategy as for the two-point function. By expressing $T$ and $\widetilde{T}$ in terms of $\epsilon$ (see \eqref{eq:TtEps} and \eqref{eq:TEps}) we get differential constraints:
\begin{equation}
\label{eq:Trels}
\begin{split}
   \langle\bar{\partial}\epsilon_1\, \bar{\partial} \epsilon_2\, \bar{\partial} \epsilon_3 \rangle
   & = \frac{72}{c^2} \frac{z_{12}z_{23}z_{13}}{\bar{z}_{12}\bar{z}_{23}\bar{z}_{13}} \,, \qquad\qquad\; \langle \partial^3 \epsilon_1\, \bar{\partial} \epsilon_2\, \bar{\partial} \epsilon_3 \rangle 
     = - \frac{72}{c^2} \frac{z_{23}^4}{z_{12}^2 \bar{z}_{23}^2 z_{13}^2}
    \\
   \langle \partial^3 \epsilon_1\, \partial^3 \epsilon_2\, \partial^3 \epsilon_3 \rangle 
   &= -\frac{12^3}{c^2} \frac{1}{z_{12}^2 z_{23}^2 z_{13}^2} \,,\qquad\;\;\,
   \langle \partial^3 \epsilon_1\, \partial^3 \epsilon_2\, \bar{\partial} \epsilon_3 \rangle 
   = \frac{12^3}{c^2}  \frac{\bar{z}_{12} z_{23} z_{13}}{z_{12}^5 \bar{z}_{23} \bar{z}_{13}} 
\end{split}
\end{equation}
as well as four more equations arising from permutations of the insertions in the correlators in the right column.
These constraints have the following solution:
\begin{equation}
\begin{split}
\langle \epsilon_1 \epsilon_2 \epsilon_3 \rangle = \langle \epsilon_1 \epsilon_2 \epsilon_3 \rangle_{\rm phys.} + 
 \langle \epsilon_1 \epsilon_2 \epsilon_3 \rangle_{\rm shad.} \\
\end{split}
\end{equation}
where the physical three-point function is the purely holomorphic quantity:
\begin{equation}\label{eq:eps3phys}
\boxed{ 
\begin{split}
\;\;\langle \epsilon_1 \epsilon_2 \epsilon_3 \rangle_{\rm phys.} = \frac{24}{c^2} \, z_{12} z_{23} z_{13} \bigg[ \left\lbrace\text{Li}_2\!\left( \frac{{z}_{13}}{{z}_{12}}\right)-\text{Li}_2\!\left( \frac{{z}_{12}}{{z}_{13}}\right)\right\rbrace& + \left\lbrace\text{Li}_2\!\left( \frac{{z}_{12}}{{z}_{32}}\right)-\text{Li}_2\!\left( \frac{{z}_{32}}{{z}_{12}}\right)  \right\rbrace\nonumber\\
& + \left\lbrace\text{Li}_2\!\left( \frac{{z}_{23}}{{z}_{13}}\right)-\text{Li}_2\!\left( \frac{{z}_{13}}{{z}_{23}}\right)   \right\rbrace\bigg]  \;\; 
\end{split}
}
\end{equation}
and the shadow piece needs to be chosen such that the full three-point function $\langle \epsilon\epsilon\epsilon\rangle$ is $(i)$ covariant and $(ii)$ consistent with \eqref{eq:Trels}. While it is easy to find expressions for $\langle \epsilon_1 \epsilon_2 \epsilon_3 \rangle_{\rm shad.}$ that satisfy  condition $(ii)$, the first condition is harder to implement. Since we will not be interested in computing shadow blocks anyway, we shall not explore these intricacies further. Note that $\langle \epsilon_1 \epsilon_2 \epsilon_3 \rangle$ is symmetric under permuting any of the labels (these permutations are captured by the permutation group $S_3$), as one would expect for a Euclidean correlation function. For instance, \eqref{eq:eps3phys} can be written as a sum over the six possible permutations of of a single term:
\begin{equation}
\begin{split}
\langle \epsilon_1 \epsilon_2 \epsilon_3 \rangle_{\rm phys.} &= \frac{24}{c^2} \, \sum_{\pi \in S_3} (z_{\pi(1)}-z_{\pi(2)}) (z_{\pi(2)}-z_{\pi(3)}) (z_{\pi(1)}-z_{\pi(3)})\; \text{Li}_2\!\left( \frac{{z}_{\pi(1)}-z_{\pi(3)}}{{z}_{\pi(1)}-z_{\pi(2)}}\right)\,.
\end{split}
\end{equation}

Again the correlator $\langle \epsilon \epsilon \epsilon \rangle$ splits into a physical and shadow piece, which are distinguished by their short distance monodromy. We could add various integration constants to the final output of this procedure, but they will not contribute in physical observables. Thus we choose them such that the vertex takes a particularly simple form. 
Now when computing the six-point star channel block, we can perform the monodromy projection at the outset by simply working with the physical vertex $\langle \epsilon \epsilon\epsilon \rangle_{\rm phys.}$ and dropping the shadow contribution.

\subsection{Explicit form of the star channel \texorpdfstring{$T$}{T} block}

We can now compute the global six-point block for pairs of identical operators. It is given by the three-point function of bilocals in \eqref{eq:G6def}, using the physical vertex function $\langle \epsilon_1 \epsilon_2 \epsilon_3 \rangle_{\rm phys.}$. Conceptually, this computation is quite simple: it is just a linear combination of the connected three-point function $\langle \epsilon_1 \epsilon_2 \epsilon_3 \rangle_{\rm phys.}$ (and its derivatives). This computation is a purely algebraic task (it does not involve any conformal integrals, nested infinite sums, or solving coupled partial differential equations). The resulting algebraic expression can be simplified and gives:
\begin{equation}\label{eq:G6res}
\boxed{
   \;\;\Gstar(z,u,v) = - \frac{144 \, h_Xh_Yh_W}{c^2}  \left[  {\cal I}\left( z,u,v \right) + {\cal I}\left( z,v,u \right) + {\cal I}\left(\frac{1}{z}, \frac{u}{z},\frac{v}{z} \right) +  {\cal I}\left(\frac{1}{z}, \frac{v}{z},\frac{u}{z}  \right) \right]
\;\; }
\end{equation}
where the function $\mathcal{I}$ in the above expression is:
\begin{equation}\label{eq:Idef}
\begin{split}
  {\cal I}(z,u,v) &\equiv 1+\frac{1}{(u-v)(1-z)} \bigg[ u \left( \frac{u(v-u+z(1-v))}{1-u} +2(z-v) \right) \log u \\
  &\qquad\qquad\qquad\qquad\quad\, - (1-u) \left( \frac{(1-u)(zv+u)}{u}+2(z-v) \right) \log (1-u) \\
  &\qquad\qquad\quad\qquad\qquad\,- 2(uv-z) \big( \text{Li}_2(u) - \text{Li}_2(1-u) \big) \bigg]
\end{split}
\end{equation}
Note the intricate branch cut structure due to the logarithms and dilogarithms. Analytic continuation to the second sheet of various cross-ratios will be crucial in the analysis of out-of-time-ordered correlation functions in section \ref{sec:chaos}. 
Compared to the well known four-point identity block, note the increased transcendentality due to the appearance of dilogarithms. We will now discuss a few consistency checks of the above result.

\paragraph{Casimir equations:} The global block \eqref{eq:G6res} satisfies the defining six-point Casimir equations, i.e., \eqref{eq:cas1}--\eqref{eq:cas3}. This is trivial to check explicitly. Our derivation in \S\ref{sec:derivation} makes this property manifest from the beginning due to the identity \eqref{eq:Casimir4ptB1} and the definition of the block in terms of reparametrization modes, \eqref{eq:G6def}.

\paragraph{OPE limits:} As identical external operators approach each other, the six-point block should reduce to a suitable five-point block involving the remaining two pairs of identical operators in addition to an external stress tensor. These OPE limits take the following form:
{\small
\begin{equation}
\label{eq:OPElimits}
\hspace{-.25cm}
\begin{split}
 \lim_{y_1 \rightarrow y_2 \equiv y}  \Gstar &\sim - \frac{8h_Xh_Yh_W}{c^2}\, \frac{(y_1-y_2)^2 (x_1-x_2)^2}{(x_2-y)^2 (w_1-y)^2} \; g^{(\text{c})}_5 \!\left( \frac{(x_1-x_2){(w_1-y)}}{(w_1-x_2)(x_1-y)} , \, \frac{(w_1-w_2)(x_2-y)}{(w_1-x_2){(w_2-y)}} \right) \\
      \lim_{x_1 \rightarrow x_2\equiv x}  \Gstar &\sim - \frac{8h_Xh_Yh_W}{c^2}\, \frac{(x_1-x_2)^2 (y_2-w_1)^2}{(y_2-x)^2 (w_2-x)^2} \; g^{(\text{c})}_5\left( \frac{(y_1-y_2){(w_1-x)}}{(w_1-y_2)(y_1-x)} , \, \frac{(w_1-w_2)(y_2-x)}{(w_1-y_2)(w_2-x)} \right)\\
   \lim_{w_1 \rightarrow w_2\equiv w}   \Gstar &\sim - \frac{8h_Xh_Yh_W}{c^2}\, \frac{(w_1-w_2)^2(y_2-x_1)^2 }{(y_2-w)^2 (x_1-w)^2} \; g^{(\text{c})}_5 \left( \frac{(y_1-y_2)(x_1-w)}{(x_1-y_2)(y_1-w)} , \, \frac{(x_1-x_2)(y_2-w)}{(x_1-y_2)(x_2-w)} \right)
\end{split}
\end{equation}
}\normalsize
where the ``bare'' five-point (comb channel) block with two pairs of identical external operators and one external stress tensor takes the following form as a function of two cross-ratios \cite{Rosenhaus:2018zqn}:\footnote{ The function $g_5^{(\text{c})}$ was was denoted as $g^{h,h,2,h',h'}_{2,2} (\chi_1,\chi_2)$ in \cite{Rosenhaus:2018zqn}.}
\begin{equation*}
g^{(\text{c})}_5 (\chi_1,\chi_2)   = \chi_1^{2} \,  \chi_2^{2}  \, F_2 ( 2,2,2;4,4;\chi_1,\chi_2)\qquad\qquad\qquad\qquad\qquad\qquad\qquad\qquad\qquad\qquad\qquad\qquad
\end{equation*}
\vspace{-.3cm}
{\footnotesize
\begin{equation}
\begin{split}
   &\;\; =  \frac{6}{\chi_1\chi_2} \bigg[ \left( 1-2\chi_2 -\chi_1^2 -\chi_1\chi_2 \right) (1-\chi_1)^2 \log(1-\chi_1) +  \left( 1-2\chi_1 -\chi_2^2 -\chi_1\chi_2 \right) (1-\chi_2)^2 \log(1-\chi_2) \\
   &\quad\;\;  - \left(1-\chi_1^2 - \chi_2^2 + \chi_1\chi_2 \right) (1-\chi_1-\chi_2)^2 \log(1-\chi_1-\chi_2) - \chi_1\chi_2  \left( 1-\chi_1 - \chi_2 + \frac{\chi_1\chi_2}{2} + \chi_1^2 + \chi_2^2  \right)  \bigg] 
\end{split}
\end{equation}
}where $F_2$ is the second Appell series, which evaluates to the explicit expression in the second line for our particular configuration of stress tensor exchanges.

Note that each line in \eqref{eq:OPElimits} contains a factor such as $(y_1-y_2)^2$ that approaches zero in the respective OPE limit $y_1 \rightarrow y_2$ with weight 2, corresponding to fusion of two external operators into a stress tensor. The remaining part of the prefactors should be understood as the ``leg factors'' used in \cite{Rosenhaus:2018zqn} in order to have a canonical normalization of the bare five-point block block $g_5^{(\text{c})}$. It is straightforward to express the coordinate dependence of the five-point blocks in \eqref{eq:OPElimits}  in terms of the two independent combinations of cross-ratios out of $(z,u,v)$ which remain finite in the respective OPE limit.\footnote{ The limits $z_{34} \rightarrow 0$ and $z_{56} \rightarrow 0$ respectively correspond to $z\rightarrow 1$ with independent parameters $(u,v)$, and $u\rightarrow v$ with independent parameters $(z,u)$. In the limit $z_{12}\rightarrow 0$ all of $z,u,v \rightarrow 1$, but one can choose independent parameters such as $\big( \frac{1-z}{z-u}, \, \frac{1-z}{z-v} \big)$ to parametrize the five-point block.} We refer the reader to \cite{Parikh:2019dvm,Jepsen:2019svc} for more details on OPE limits of higher-point blocks.

 \begin{figure}
	\centering     
	\includegraphics[width=.25\textwidth]{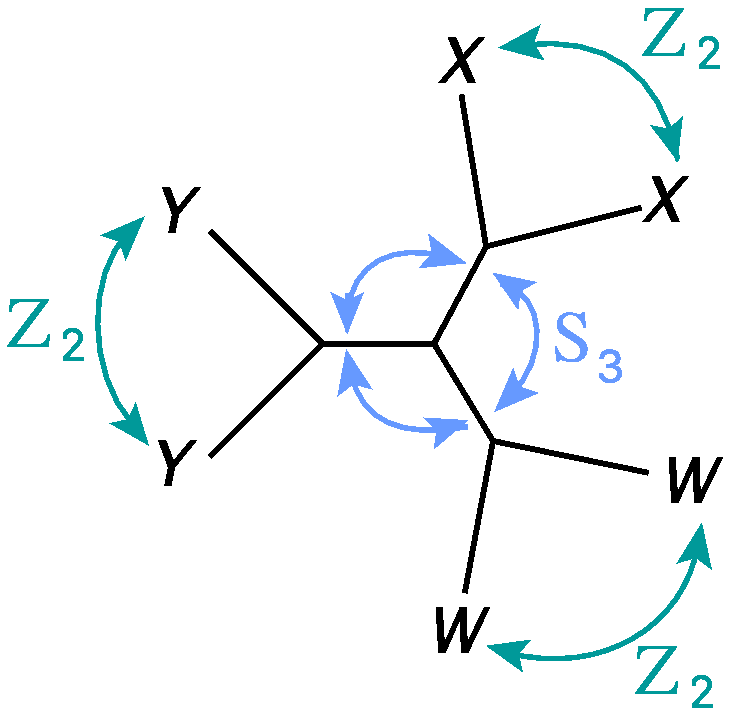}
	\caption{Permutation symmetries of the star channel identity block.}
	\label{fig:symms}
\end{figure}

\paragraph{Symmetric presentation of the star channel block:}
The star channel block is invariant under a number of permutation symmetries of the form $\big(\mathbb{Z}_2^{(X)} \times \mathbb{Z}_2^{(Y)} \times \mathbb{Z}_2^{(W)}\big) \rtimes  S_3$:
\begin{itemize}
\item From the definition \eqref{eq:G6def}, it is clear that $\Gstar$ should have a $\mathbb{Z}_2^{(X)}\times \mathbb{Z}_2^{(Y)} \times \mathbb{Z}_2^{(W)}$ symmetry corresponding to exchanging any of the pairs $X_1 \leftrightarrow X_2$, $Y_1 \leftrightarrow Y_2$, and $W_1 \leftrightarrow W_2$.
\item From the three-point vertex $\langle \epsilon_1 \epsilon_2 \epsilon_3 \rangle_{\rm phys.}$ it is manifest that $\Gstar$ should further have an $S_3$ symmetry corresponding to all permutations of the pairs $(X_1,X_2)$, $(Y_1,Y_2)$, $(W_1,W_2)$. This corresponds to performing any of the following exchanges:
{\small
\begin{equation}
  x_i \leftrightarrow y_i \,,\quad\;\;
  y_i \leftrightarrow w_i \,,\quad\;\;
  w_i \leftrightarrow x_i \,,\quad\;\;
  (x_i,y_i,w_i) \leftrightarrow (y_i,w_i,x_i) \,,\quad\;\;
  (x_i,y_i,w_i) \leftrightarrow (w_i,x_i,y_i) \,.
\end{equation}}
Since the $S_3$ permutations do not commute with the $\mathbb{Z}_2$ symmetries when acting on the space of operator insertion points, the full permutation group is a semi-direct product.\footnote{ We thank J.-F.\ Fortin, W.-J.\ Ma, and W.\ Skiba for bringing this subtlety to our attention.}

\end{itemize}
Figure \ref{fig:symms} illustrates the symmetries pictorially.  These symmetries of the identity  block were manifest in our derivation in \S\ref{sec:derivation}.
However, not all of the symmetries of the star channel block are manifest in \eqref{eq:G6res}.
The full result for the star channel block indeed takes the following form:
\begin{equation}
\label{eq:G6symm}
\begin{split}
 \Gstar &= - \frac{12 h_Xh_Yh_Z}{c^2} \sum_{\sigma_x \in \mathbb{Z}_2} \sum_{\sigma_y \in \mathbb{Z}_2} \sum_{\sigma_w \in \mathbb{Z}_2} \sum_{\pi \in S_3} \; {\cal I}\left( z_{\sigma_x,\sigma_y,\sigma_w,\pi},\;u_{\sigma_x,\sigma_y,\sigma_w,\pi},\;v_{\sigma_x,\sigma_y,\sigma_w,\pi} \right)
\end{split}
\end{equation}
where the function ${\cal I}(z,u,v)$ was given in \eqref{eq:Idef}, and $z_{\sigma_x,\sigma_y,\sigma_w,\pi}$ etc.\ denote the cross-ratios evaluated on insertion points that are permuted according to the composition of the four permutations $\sigma_x \circ \sigma_y \circ \sigma_w \circ \pi$. On the cross-ratios these permutations act explicitly as follows:
\small{
\begin{equation}
\label{eq:perms}
\begin{split}
  \sigma_x: \quad & (z,u,v)  \mapsto  (z',u',v') \in \left\{ (z,u,v) , \; \left( \frac{1}{z},\frac{u}{z},\frac{v}{z} \right) \right\}\,,\\
   \sigma_y: \quad & (z,u,v)  \mapsto  (z',u',v') \in \left\{ (z,u,v) , \;\left( \frac{1}{z},\frac{1}{u},\frac{1}{v} \right)\right\} \,,\\
  \sigma_w: \quad & (z,u,v)  \mapsto  (z',u',v') \in \Big\{ (z,u,v) , \;\left( z,v,u \right)  \Big\}\,,\\
   \pi: \quad & (z,u,v)  \mapsto (z',u',v') \in \bigg\{ (z,u,v) ,\; \left( z , \frac{z-u}{1-u},\frac{z-v}{1-v} \right) ,\; \left(\frac{(1-u)(z-v)}{(1-v)(z-u)} ,\frac{1-u}{1-v},\frac{v(1-u)}{u(1-v)}\right), \\
   &\qquad\qquad\qquad\qquad\qquad \left( \frac{v}{u} ,\frac{1-v}{1-u} , \frac{z-v}{z-u} \right)  ,\;\left( \frac{v}{u},\frac{1}{u},\frac{z}{u} \right)  , \; \left( \frac{(1-u)(z-v)}{(1-v)(z-u)}, \frac{1-u}{z-u} , \frac{z(1-u)}{z-u} \right) \bigg\}
\end{split}
\end{equation}
}\normalsize
The expression \eqref{eq:G6symm} of writing the block makes manifest all of its symmetries. However, for practical purposes, it is often useful to simplify the 48 terms in \eqref{eq:G6symm} at the expense of some symmetries still being present but not being manifest anymore. The result is \eqref{eq:G6res}.

\paragraph{A remark on the coefficient:} It goes without saying that the global block, being a solutions of a particular set of differential equations, is defined up to an overall normalization. Yet here we have written it with a very particular prefactor, including explicit dependence on external dimensions and the central charge $c$. The reason for this is that we view it as contributing individually to a large-$c$ expansion to the entire \emph{Virasoro} identity block. The block presented in this section will come with these particular coefficients when appearing in the said expansion of the Virasoro block, as dictated by the Feynman rules for the reparametrization modes.

\section{The star-channel Virasoro identity block }
\label{sec:virasoro}

The shadow operator formalism is only appropriate for global blocks. However, we have intentionally presented it in a way that allows an immediate generalization to the Virasoro case. We will argue that it is sensible to compute higher order exchanges of reparametrization modes between bilocal operators. From the previous discussion it is already clear one should think of self-interactions of $\epsilon$ as originating from stress tensor correlators. We will now explain how to study the six-point {\it Virasoro} block in the star channel for certain ranges of operator dimensions.

In order to study Virasoro blocks, we will interpret the $\epsilon$ field in a way that differs from a simple rewriting of the shadow stress tensor. We will instead think of it as a reparametrization field whose dynamics captures arbitrary multi-$T$ exchanges. For global blocks the relation between reparametrization modes and the shadow of $T$ was clarified in \cite{Haehl:2019eae}. It was furthermore verified in \cite{Cotler:2018zff,Jensen:2019cmr} that the nonlinear extension of the formalism is appropriate for highly nontrivial aspects of Virasoro blocks.

Inspired by these studies, our proposal is that the stress tensor contributions to the star channel Virasoro six-point block of pairwise equal external operators is given by a normalized sum over all reparametrization mode exchanges:\footnote{ This expression should be thought of as $\frac{\langle {\cal B}_Y{\cal B}_W{\cal B}_X \rangle}{\langle {\cal B}_Y \rangle \langle {\cal B}_W \rangle \langle {\cal B}_X \rangle} \left( \frac{\langle {\cal B}_X{\cal B}_Y\rangle}{\langle {\cal B}_X\rangle\langle{\cal B}_Y\rangle}  \frac{\langle {\cal B}_Y{\cal B}_W\rangle}{\langle {\cal B}_Y\rangle\langle{\cal B}_W\rangle}  \frac{\langle {\cal B}_X{\cal B}_W\rangle}{\langle {\cal B}_X\rangle\langle{\cal B}_W\rangle} \right)^{-1}$. This particular normalization is also motivated by the expectation that the Virasoro blocks should exponentiate: if the four- and six-point blocks do exponentiate, then dividing blocks in this way serves to subtract disconnected (lower-point) contributions in the exponential. One is left with the genuinely connected six-point piece of the block. As we will see, this indeed works as advertised.}
{\small
\begin{equation}
\label{eq:V6def}
\begin{split}
   &{\cal V}^{(6,\,\text{star})}_{T} = \frac{\big{\langle} {\cal B}_{\epsilon,h_X}(x_1,x_2)\, {\cal B}_{\epsilon,h_Y}(y_1,y_2)\,  {\cal B}_{\epsilon,h_W}(w_1,w_2) \big{\rangle} \; \big{\langle} {\cal B}_{\epsilon,h_X}(x_1,x_2)\big{\rangle}\;\big{\langle} {\cal B}_{\epsilon,h_Y}(y_1,y_2)\big{\rangle}\;\big{\langle}  {\cal B}_{\epsilon,h_W}(w_1,w_2) \big{\rangle} }{ \big{\langle} {\cal B}_{\epsilon,h_X}(x_1,x_2)\, {\cal B}_{\epsilon,h_Y}(y_1,y_2) \big{\rangle}\; \big{\langle} {\cal B}_{\epsilon,h_Y}(y_1,y_2)\, {\cal B}_{\epsilon,h_W}(w_1,w_2) \big{\rangle}\; \big{\langle} {\cal B}_{\epsilon,h_X}(x_1,x_2)\, {\cal B}_{\epsilon,h_W}(w_1,w_2) \big{\rangle}} \bigg{|}_{\rm phys.}
   \end{split}
\end{equation}
}\normalsize
Here, ${\cal B}_{\epsilon,h}$ denotes the all-order bilocal operator, describing the coupling of pairs of primaries to any number of $\epsilon$. To define this coupling, it is useful to take the perspective that $\epsilon$ describes the universal  reparametrizations $z \rightarrow f(z,\bar{z}) = z + \epsilon(z,\bar{z}) + {\cal O}(\epsilon^2)$. The higher order bilocals can then be read off from a reparametrized conformal two-point function:
\begin{equation}
\label{eq:verticesEucl}
\begin{split}
{\cal B}_{\epsilon,h}(z_1,z_2) &\equiv (z_1-z_2)^{2h}  \left(  \frac{\partial f(z_1,\bar{z}_1) \, \partial f(z_2,\bar{z}_2)}{(f(z_1,\bar{z}_1)-f(z_2,\bar{z}_2) )^2 } \right)^{h}  = 1 +  \sum_{p\geq 1}  {\cal B}_{\epsilon,h}^{(p)}(z_1,z_2)\,,
 \end{split}
\end{equation}
where ${\cal B}_{\epsilon,h}^{(p)}$ is the term that occurs at $p$-th order in $\epsilon$.
In order to expand to higher orders in $\epsilon$, we need to decide how to expand the function $f(z) = z + \epsilon + \ldots$. The natural choice turns out to be $f(z) = e^{\epsilon\partial} z = z + \epsilon + \frac{1}{2} \, \epsilon \partial \epsilon + \ldots$. This exponentiation simply describes the finite action of an infinitesimal diffeomorphism. The bilocals then have a nice structure:\footnote{ We thank W.\ Reeves and M.\ Rozali for discussions about this.} ${\cal B}_{\epsilon,h}^{(p)}$ can be built out of ``atomic'' building blocks $b^{(q)}$, which are the truly connected pieces that everything else is built from. The atomic bilocal $b^{(p)}$ shows up for the first time as the ${\cal O}(h)$ term in ${\cal B}_{\epsilon,h}^{(p)}$:
\begin{equation}
\begin{split}
  {\cal B}_{\epsilon,h}^{(1)}(z_1,z_2) &= b^{(1)}(z_1,z_2) \,, \\
    {\cal B}_{\epsilon,h}^{(2)}(z_1,z_2) &= \frac{1}{2!} \left(b^{(1)}(z_1,z_2)\right)^2 + b^{(2)}(z_1,z_2) \,, \\
    {\cal B}_{\epsilon,h}^{(3)}(z_1,z_2) &= \frac{1}{3!} \left(b^{(1)}(z_1,z_2)\right)^3 + b^{(2)}(z_1,z_2) \, b^{(1)}(z_1,z_2) +  b^{(3)}(z_1,z_2) \,, \quad \ldots 
\end{split}
\end{equation}
where the first few atomic bilocals are given by the following kinematic space fields:
\begin{equation}
\label{eq:verticesEucl2}
\begin{split}
b^{(1)}(z_1,z_2)  &= \;
\begin{gathered} \includegraphics[width=.06\textwidth]{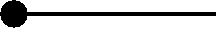} \end{gathered}  \;
=
h \left[ \partial \epsilon_1 +\partial \epsilon_2  -2\, \frac{\epsilon_1 - \epsilon_2}{z_1-z_2} \right]\\
b^{(2)}(z_1,z_2) &= \;
\begin{gathered}\vspace{-.2cm} \includegraphics[width=.06\textwidth]{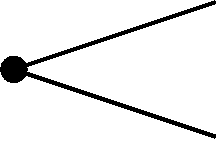} \end{gathered}  \;
=
h \left[ \frac{1}{2} \left( \epsilon_1 \partial^2 \epsilon_1 + \epsilon_2 \partial^2 \epsilon_2 \right) - \frac{1}{z_{12}} \left( \epsilon_1 \partial \epsilon_1 - \epsilon_2 \partial \epsilon_2 \right) + \frac{1}{z_{12}^2} (\epsilon_1-\epsilon_2)^2 \right] \\
 b^{(3)}(z_1,z_2) &= \;
\begin{gathered}\vspace{-.2cm} \includegraphics[width=.06\textwidth]{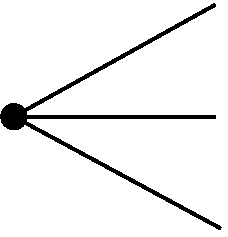} \end{gathered}  \;
= h \left[ \frac{z_{12}^2}{6} \, \epsilon_1 \, \partial_1 \left( \frac{\epsilon_1 \partial^2 \epsilon_1}{z_{12}^2}\right)+   \frac{\epsilon_1(\epsilon_1-\epsilon_2) \partial \epsilon_1}{z_{12}^2} -  \frac{\epsilon_1(\partial \epsilon_1)^2}{3z_{12}} - \frac{(\epsilon_1-\epsilon_2)^3}{3z_{12}^3} + (z_1 \leftrightarrow z_2)\right]
\end{split}
\end{equation}
Diagrammatically, we can write the general bilocal at order $p$ as follows:
\begin{equation}
\label{eq:BpAtomic}
\begin{split}
 {\cal B}_{\epsilon,h}^{(p)}(z_1,z_2) &= \sum_{\substack{ \{a_1,\ldots,a_p\}:  \\  \sum k a_k = p}}  \;\; \frac{1}{a_1! \cdots a_p!} \;\;
\begin{gathered}\vspace{-.2cm} \includegraphics[width=.4\textwidth]{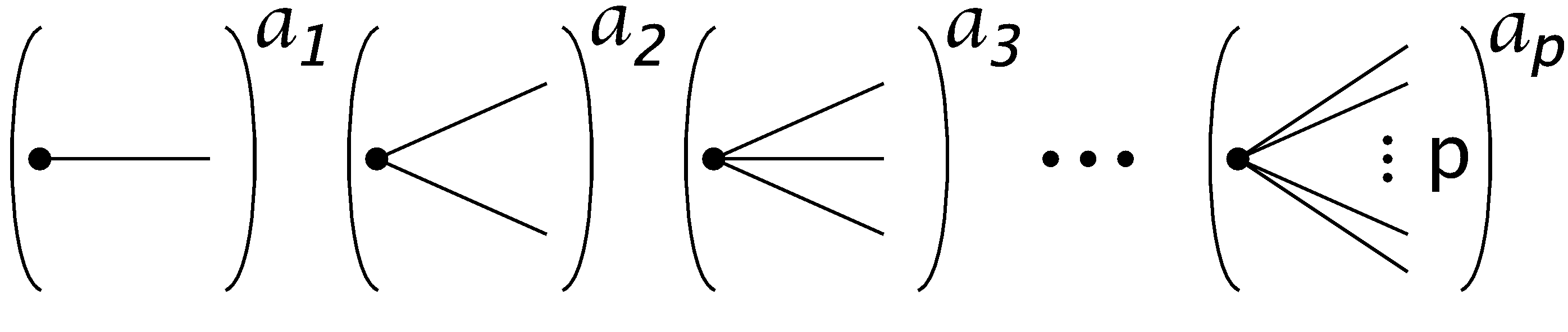} \end{gathered} 
\end{split}
\end{equation}
where the inner sum runs over integer partitions $p=1 \cdot a_1 + 2 \cdot a_2 + \ldots +p \cdot  a_p$ and the power in $h$ of any given term is $q \equiv {\sum a_k}$.
This expression is just a sum over collections of atomic pieces such that each collection has a total of $q$ dots (i.e., powers of $h$) and $p$ emanating lines (i.e., reparametrization fields $\epsilon$). More simply, we can observe that the symmetry factors conspire to give an exponential structure: 
\begin{equation}
\label{eq:BpAtomic2}
\boxed{
\;\; {\cal B}_{\epsilon,h}(z_1,z_2)  = \exp \left[  \sum\nolimits_{q\geq1} b^{(q)}(z_1,z_2) \right] = \exp \bigg(  \;\begin{gathered}\vspace{-.2cm} \includegraphics[width=.3\textwidth]{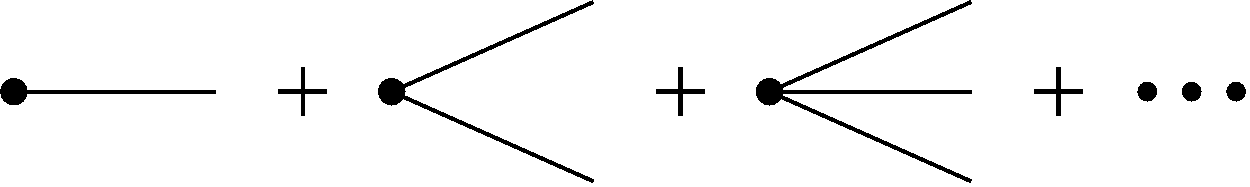} \end{gathered} \;\;\bigg)\;\;
 }
\end{equation}
as one can easily check from the above definitions and properties. Note that the expression inside the exponent is ${\cal O}(h)$.

\paragraph{Virasoro block as a sum of diagrams:} The expression \eqref{eq:V6def} corresponds to fusing pairs of external operators into any number of stress tensors and descendants. By expanding the bilocals to arbitrary orders in $\epsilon$, it is clear that the Virasoro block \eqref{eq:V6def} corresponds to an infinite sum of reparametrization mode Feynman-type diagrams: each bilocal contains sums over collections of atomic graphs such as \eqref{eq:BpAtomic}. Inside a correlation function these get Wick contracted in all possible ways (i.e., all open lines need to be connected up to form $\epsilon$-propagators or self-interaction vertices). {\it The normalization in \eqref{eq:V6def} is engineered such as to precisely remove all disconnected diagrams;} here, by `disconnected' we mean any diagram with at least one contraction of atomic bilocals that involves not all three of the operators $X,Y,W$. We are left with an infinite sum of products of connected diagrams. This is perhaps not immediately clear, but will become more transparent in our examples below.

In general we can note that any propagator scales as $c^{-1}$, every self-interaction vertex scales as $c$, and any $p$-th order external bilocal operator ${\cal B}_{\epsilon,h}^{(p)}$ has terms of orders $h^q$ for $q=1,\ldots,p$. Under certain assumptions about the scaling of operator dimensions with central charge, a subset of diagrams dominates at large $c$, which we can sum explicitly.  We shall now illustrate this for simple cases.

\subsection{Light external operators: \texorpdfstring{$h_i \sim {\cal O}(c^0)$}{c0}}
\label{subsec:light}

We refer to the regime where all operator dimensions scale as ${\cal O}(c^0)$ as `semiclassical'. In terms of reparametrization mode diagrams we are interested in diagrams with few propagators and vertices in order to get leading results for large $c$. The leading connected diagrams (for the holomorphic part of the Virasoro block) are of the following form:
\begin{equation}
\begin{split}\label{eq:virdiags}
   {\cal V}_T^{(6,\,\text{star})} 
      &=1+\Big\{ \left\langle b^{(1)}_X\,b^{(1)}_Y\,b^{(1)}_W \right\rangle + \left[  \left\langle b^{(1)}_X \, b^{(1)}_Y \, b^{(2)}_W \right\rangle- \left\langle b^{(1)}_X\, b^{(1)}_Y \right\rangle \left\langle b^{(2)}_W \right\rangle   +  (W \leftrightarrow X) + (W \leftrightarrow Y)  \right] \Big\}  + \ldots\\
      &= 1 + \;\;
   \begin{gathered} \vspace{-.1cm}\includegraphics[width=.45\textwidth]{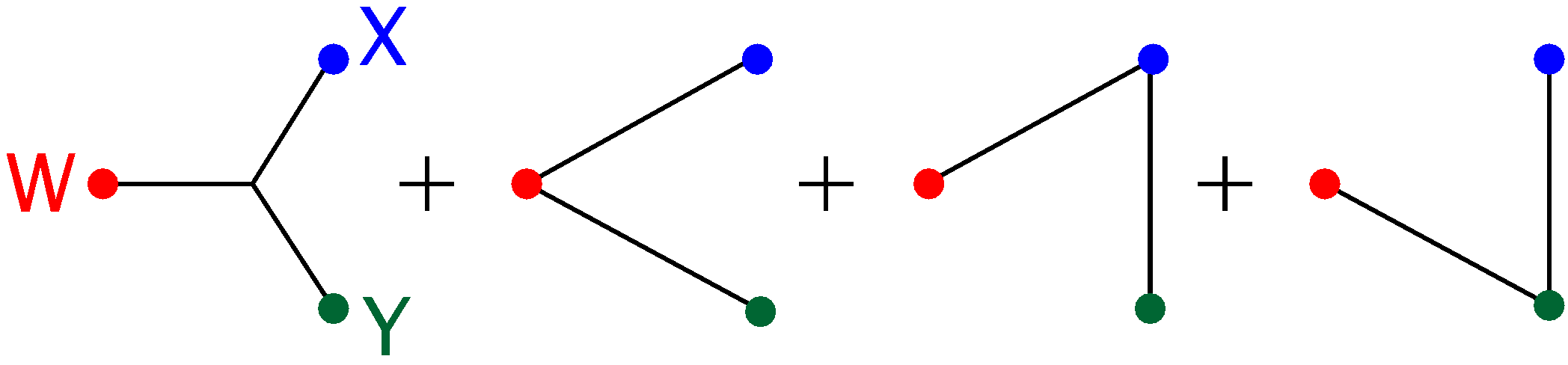} \end{gathered} \;+ \ldots 
\end{split}
\end{equation}
where we abbreviate $b^{(p)}_X \equiv b^{(p)}(x_1,x_2)$ etc., and colors are merely to distinguish the operators graphically. Note that these leading diagrams are indeed fully connected. The square bracket has various four-point functions of $\epsilon$. The leading contribution will be the disconnected Wick contractions (of the form $\langle \epsilon_X\epsilon_Y\epsilon_W\epsilon_W\rangle \rightarrow 2 \langle\epsilon_X\epsilon_W\rangle\langle\epsilon_Y\epsilon_W\rangle + \langle\epsilon_X\epsilon_Y\rangle\langle\epsilon_W\epsilon_W\rangle$, and our normalization explicitly removes the second term as it is a truly a part of a four-point block). Diagrammatically, we capture this process compactly by writing:
\begin{equation}
\left\langle b^{(1)}_X \, b^{(1)}_Y \, b^{(2)}_W \right\rangle- \left\langle b^{(1)}_X\, b^{(1)}_Y \right\rangle \left\langle b^{(2)}_W \right\rangle
=\;
\begin{gathered} \vspace{-.2cm}\includegraphics[width=.35\textwidth]{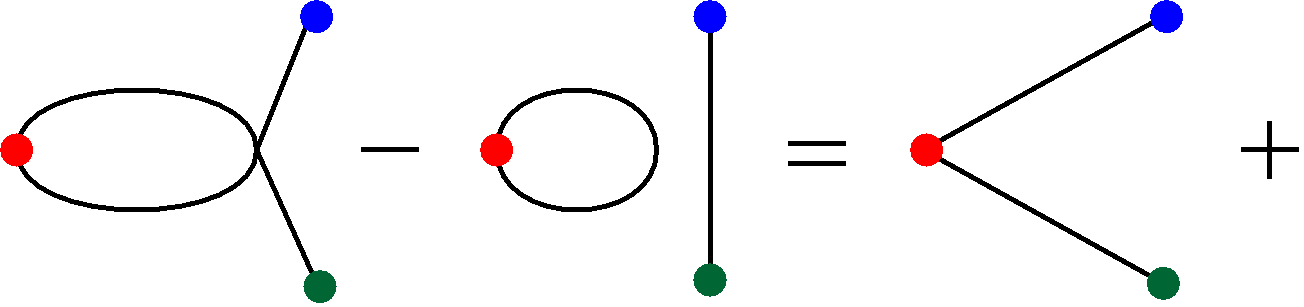} \end{gathered} \;\; {\cal O}(c^{-1})
\end{equation}
That is, the diagram on the right hand side describes the disconnected part (in the sense of Wick contractions) of the four-point function, which is nevertheless connected in a diagrammatic sense (i.e., not constructible out of blocks involving less than all of the operators).
Evaluating the expression \eqref{eq:virdiags} gives:
\begin{equation}
\label{eq:V6Result}
\boxed{ \;\;
  \text{For} \quad  h_X \sim h_Y \sim h_W \sim {\cal O}(c^{0}): \qquad
    {\cal V}_T^{(6,\,\text{star})} = 1 + \Gstar+\Gext+\mathcal{O}(c^{-3})}
\end{equation}
with the global block $\Gstar$ given in \eqref{eq:G6res}, and 
\begin{equation}
	\Gext\equiv  \frac{18 h_X h_Y h_W}{c^2} \left[ \widetilde{\cal I}(z,u,v) + \widetilde{\cal I}(z,v,u) + \widetilde{\cal I} \left( \frac{1}{z},\frac{u}{z}, \frac{v}{z} \right)   + \widetilde{\cal I} \left( \frac{1}{z},\frac{v}{z}, \frac{u}{z} \right)\right]
\end{equation}
with
{\small
\begin{equation}
\begin{split}
\widetilde{\cal I}(z,u,v)
&=  \left[ \frac{2(2+u+v)}{1-z}-\frac{1+2u^2}{1-u} - \frac{1}{1-v}  - \frac{(u-v)z}{(z-u)(z-v)} - \frac{2u(v+(2+u)z)}{z(u-v)} - \frac{8uv\, \log z}{(u-v)(1-z)}  \right]\log u \\
&\quad - \frac{4(1-u)}{(u-v)(1-z)} \left[ 1+\frac{ vz-u^2}{u}-vz+2(z-v)   \right. \\
&\qquad\qquad\qquad\qquad\quad\;\;\, \left. + \frac{4(1-v)z}{1-z} \log z + \frac{2(uv-z)}{(1-u)} \log u- \frac{4(z-u)v}{u-v} \, \log \frac{u}{v} \right] \log(1-u)
\end{split}
\end{equation}
}\normalsize
As before, we are writing these terms in a way that only makes the symmetries $\mathbb{Z}_2^{(X)} \times \mathbb{Z}_2^{(W)}$ manifest. The remaining permutation symmetries are still present but not manifest.

We have thus shown that in this particular limit the Virasoro identity block \eqref{eq:V6def} at leading nontrivial order consists of two contributions: $(i)$ the star channel global block discussed previously, and $(ii)$ another piece at ${\cal O}\big( \frac{h_X h_Y h_W}{c^2}\big)$ involving logarithms and products of logarithms. Note that both of these pieces are ${\cal O}(c^{-2})$ in the regime of operator dimensions under consideration. Any contributions at more dominant orders is not truly connected and is cancelled by the normalization in \eqref{eq:V6def}. Subleading corrections to the above expressions are ${\cal O}(c^{-3})$

We will show in section \ref{sec:chaos} that the particular piece $\Gext$ is the crucial contribution responsible for six-point scrambling.

\subsection{`Hefty' operator regime: \texorpdfstring{$h_i \sim {\cal O}(c^{1/2})$}{conehalf}}

\begin{figure}[t]
	\centering     
	\includegraphics[width=.27\textwidth]{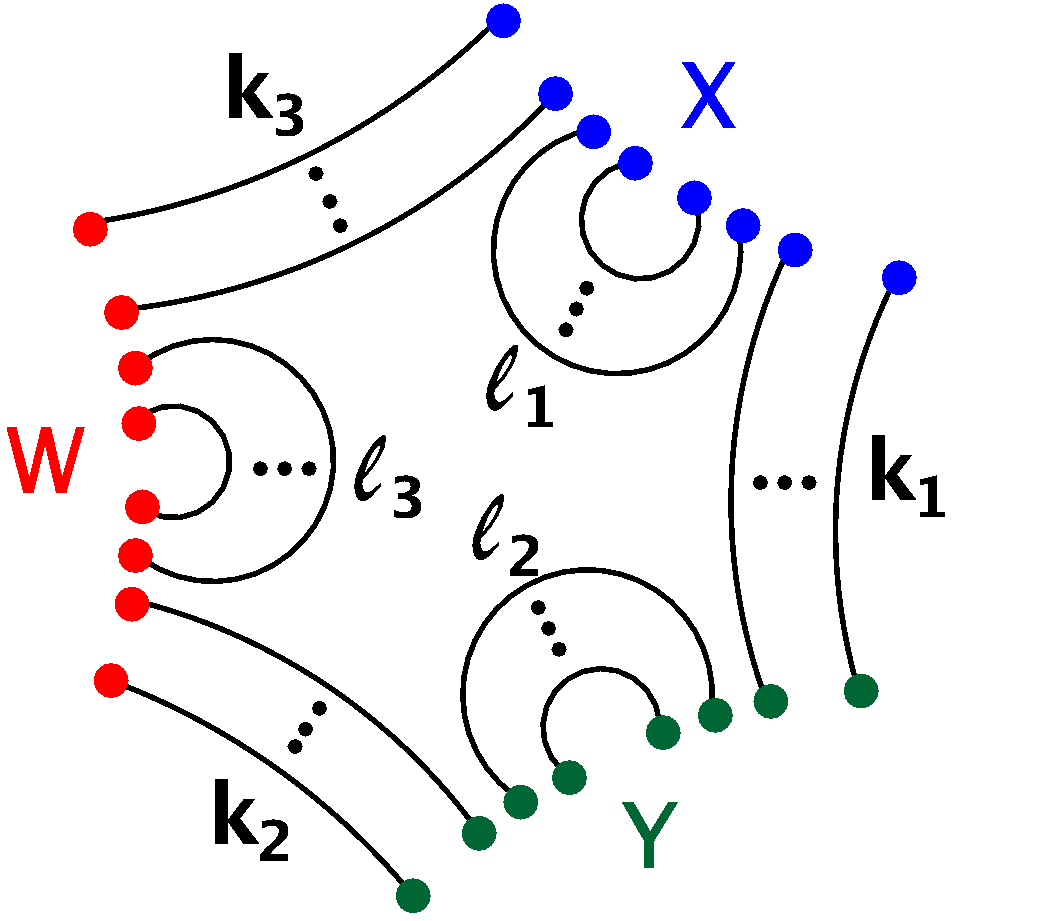}
	\caption{The leading reparametrization mode exchanges in the six-point Virasoro identity block at large $c$ in the `hefty' regime where $h_X \sim h_Y \sim h_W \sim {\cal O}(c^{1/2})$. All diagrams of this form contribute at ${\cal O}(c^0)$.}
	\label{fig:kl}
\end{figure}

If $h_X \sim h_Y \sim h_W \sim {\cal O}(c^{1/2})$, the reader can convince themselves that the leading diagrams at large $c$ are of the form illustrated in figure \ref{fig:kl}: they involve any number of propagators between any of the bilocals, but no self-interaction vertices. However, we claimed that the normalization in \eqref{eq:V6def} serves to subtract out any disconnected diagrams. If this claim is true, then any diagram of the type shown in figure \ref{fig:kl} should drop out. Confirming this presents a good check of the statement that normalization by four-point blocks corresponds to removing disconnected diagrams. 

It is in fact a matter of simple combinatorics to sum all diagrams of this form. We defer details to Appendix \ref{app:virasoro}. The result confirms that all these diagrams should be thought of as disconnected and that they do indeed cancel:
\begin{equation}
\label{eq:VirResHefty}
   \text{For} \quad  h_X \sim h_Y \sim h_W \sim {\cal O}(c^{1/2}): \qquad   {\cal V}^{(6,\,\text{star})}_{T} = 1 + {\cal O}(c^{-1/2})\,.
\end{equation}
This means that to leading order the `hefty' Virasoro block is just built out of four-point blocks, which are known \cite{Fitzpatrick:2014vua} to be exponentials of the single-$\epsilon$ exchange (see \eqref{eq:B1B1exponential}). The latter are cancelled out by the normalization in the definition \eqref{eq:V6def}. 
In a dual gravitational description, we are led to conclude that a six-point function of probes which are `hefty' in the sense of $h_i \sim {\cal O}(c^{1/2})$ does not probe gravitational self-interactions at leading order. To see nonlinear effects in gravity (even at tree level), one needs to compute subleading corrections.\footnote{ We thank A.\ Streicher for discussions about this.} 

\paragraph{Subleading contributions at ${\cal O}(c^{-1/2})$:}
We can ask what are the first subleading corrections, which contribute at ${\cal O}(c^{-1/2})$ to \eqref{eq:VirResHefty}. Diagrammatically, these come from the diagrams shown in figure \ref{fig:kl} where either: any single pair of atomic $\big(b^{(1)}_i\big)^2$ is replaced by a connected $b^{(2)}_i$, or: any three $b^{(1)}_{X}b^{(1)}_Yb^{(1)}_W$ are contracted in a three-point vertex. A trivial modification of the calculation in Appendix \ref{app:virasoro} shows that all dressings by arbitrarily many ``rungs'' $\langle b^{(1)}_{i} b^{(1)}_{j} \rangle$ and ``melons'' $\langle \big(b^{(1)}_{i}\big)^2 \rangle$ cancels out and the remaining diagrams are precisely the same as in \eqref{eq:virdiags}. In other words, the `semiclassical' result \eqref{eq:V6Result} applies in the present `hefty' regime as well:
\begin{equation}
\label{eq:VirRes}
\begin{split}
 \text{For} \quad  h_X \sim h_Y \sim h_W \sim {\cal O}(c^{1/2}): \qquad
   {\cal V}_T^{(6,\,\text{star})} &=  1 + \Gstar+\Gext+ {\cal O}(c^{-1})~.
\end{split}
\end{equation}
The only difference is that this is now an ${\cal O}(c^{-1/2})$ contribution and the next order corrections are ${\cal O}(c^{-1})$.

\subsection{`Heftier' operator regime: \texorpdfstring{$h_i \sim {\cal O}(c^{2/3})$}{conehalf}}

We can ask if there is a regime of operator dimensions for which the Virasoro identity block exponentiates in analogy with the four-point case \cite{Fitzpatrick:2014vua}. We suggest that a simple regime allowing for an analogous argument should involve `heftier' operators with $h_X \sim h_Y \sim h_W \sim c^{2/3}$. With such a scaling, the leading connected diagrams discussed in \eqref{eq:virdiags} are ${\cal O}(c^0)$. Therefore, products of such diagrams are still connected and of leading order. All of the following diagrams are ${\cal O}\big(\big(\frac{h_Xh_Yh_W}{c^2}\big)^k\big) \sim {\cal O}(c^0)$ and should be summed at leading order:
\begin{equation*}
 {\cal O}(c^0): \qquad \begin{gathered}\vspace{-.1cm} \includegraphics[width=.85\textwidth]{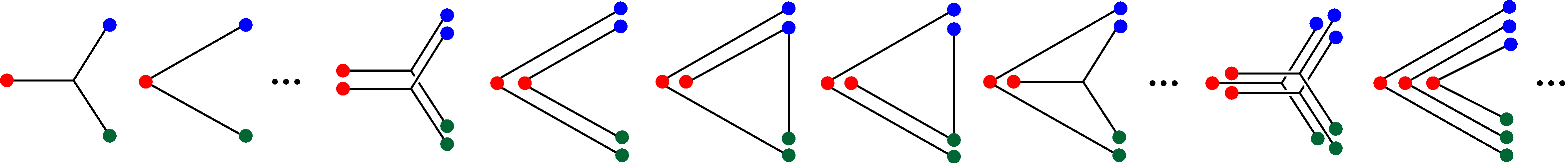}\end{gathered}
\end{equation*}
One can show that diagrams built out of products of the basic three-point ``star diagram" and the ``triangle diagram"  are in fact {\it all} connected diagrams at this order. This is easy to prove, but requires too much notation to be very illuminating. We therefore simply present a few examples of graphs which are representative of  connected diagrams that are {\it not} of leading order at large $c$:
\begin{equation*}
 {\cal O}(c^{-1/3}): \qquad \begin{gathered}\vspace{-.1cm} \includegraphics[width=.48\textwidth]{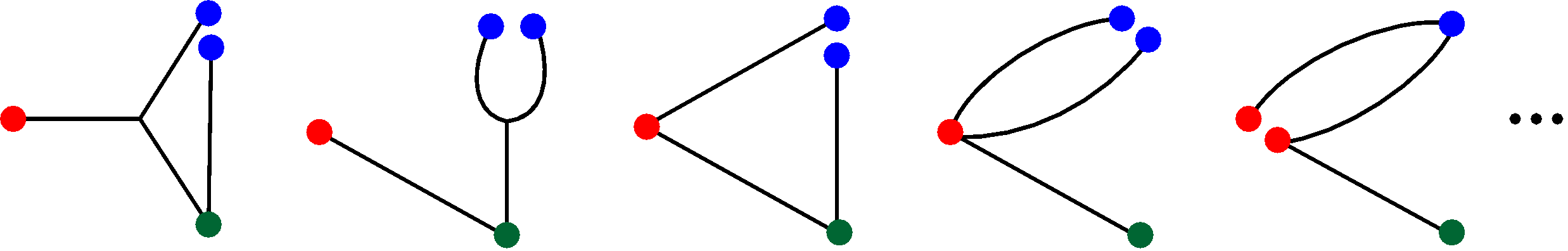}\end{gathered}
\end{equation*}
For instance, the sum of the first two diagrams combined is a shorthand for the genuinely six-point piece $\langle (b^{(1)}_X)^2 \, b^{(2)}_Y \, b^{(1)}_W \rangle -  \langle (b^{(1)}_X)^2 \rangle \langle b^{(2)}_Y \, b^{(1)}_W \rangle -\langle (b^{(1)}_X)^2  \, b^{(1)}_W \rangle \langle b^{(2)}_Y \rangle - 2 \langle b_X^{(1)}b_Y^{(2)}\rangle \langle b_X^{(1)} b_W^{(1)} \rangle + {\cal O}(c^{-4/3})$.

We can therefore proceed with summing the types of diagrams occurring at ${\cal O}(c^0)$. Simple counting arguments show that the symmetry factors work out to give precisely the desired exponentiation:
\begin{equation}
\label{eq:VirResHeftier}
   \text{For} \quad  h_X \sim h_Y \sim h_W \sim {\cal O}(c^{2/3}): \qquad   {\cal V}^{(6,\,\text{star})}_{T} =  \exp \left[  \Gstar+\Gext \right] + {\cal O}(c^{-1/3}) \,.
\end{equation} 
It thus seems that we have identified a regime of operator dimensions, which allows for a simple exponentiation argument in the six-point identity block. However, note the following subtlety: at least in terms of diagrammatic perturbation theory, the four-point blocks of `heftier' operators are not easily understood. In this case, our normalization in \eqref{eq:V6def} therefore does something very nontrivial: it removes from the six-point block all dependence on four-point blocks, the latter of which we do not understand per se. It is nevertheless intriguing that there is a simple structure in the genuinely six-point part of the block, which can be unearthed without having to study the four-point blocks themselves. We leave it for the future to understand the implications of this better.

\section{Six-point out-of-time-order correlators from the identity block}
\label{sec:chaos}

We have now collected all of the ingredients needed to calculate the identity block contributions to the six-point out-of-time-order correlator (OTOC) of scalar operators. We will compute both the contribution from the global star channel block, and from the leading order Virasoro block. Interestingly, we will find that only the latter is relevant for six-point scrambling.

Since we would like to compute the `fine-grained' chaos exponent associated with the fully connected six-point function, we must normalize by partially disconnected correlation functions. That is we are interested in computing the following object: 
\begin{align}
 	\text{OTOC}_{6pt}&\equiv\frac{\langle XYXWYW \rangle_\beta\langle XX\rangle_\beta\langle YY\rangle_\beta\langle WW\rangle_\beta}{{\langle XYXY\rangle_\beta}{\langle XXWW\rangle_\beta}{\langle YWYW \rangle_\beta}}\approx \mathcal{V}_0\,\overline{\mathcal{V}}_0+\dots\nonumber\\
 	&\approx\left(1+\Gstar+\Gext+\dots\right)\left(1+\Gbstar+\Gbarext+\dots\right)+\dots
 \end{align} 
 where we have indicated the relevant time ordering by the order of operators in the correlation function, with operators inserted later in the Lorentzian time evolution placed further to the left. In the above expression $\mathcal{V}_0$ is the identity Virasoro block, which we further decompose according to our findings in the previous section. The subscript $\beta$ indicates that we evaluate the blocks in a thermal state.
 
 Note that there is a larger space of six-point OTOCs \cite{Haehl:2017qfl,Haehl:2017eob}, of which the configuration above only presents a special case. However, it was argued in \cite{Haehl:2017pak} that the configuration above is the most interesting representative, being both {\it maximally out-of-time-order} as well as {\it maximally braided} in Euclidean time. Physically, it is distinguished by the fact that its exponential growth lasts for the longest time out of all possible inequivalent six-point OTOCs. We will confirm this expectation below.

\begin{figure}
	\centering     
	\includegraphics[width=.45\textwidth]{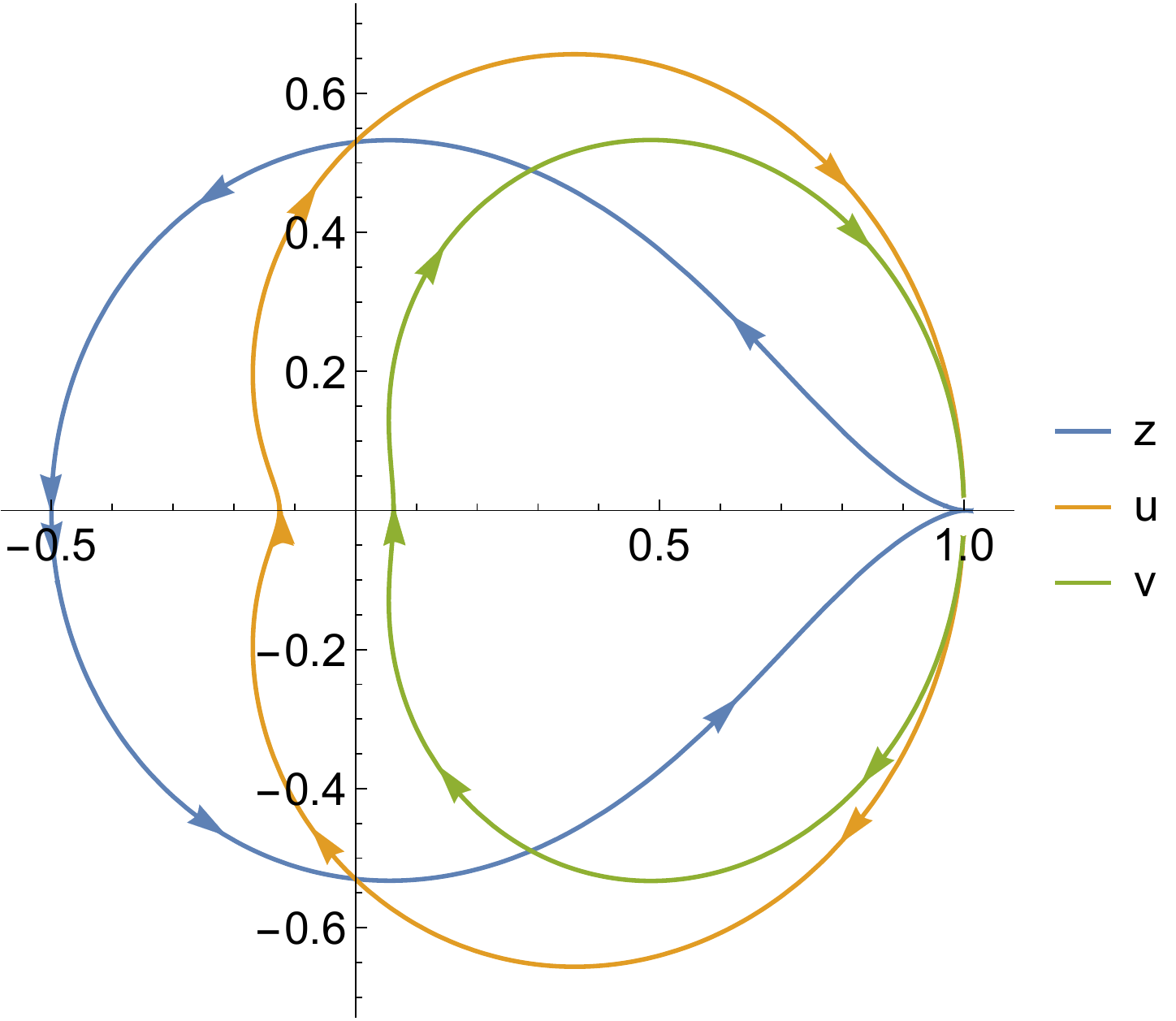}$\qquad\quad$\includegraphics[width=.45\textwidth]{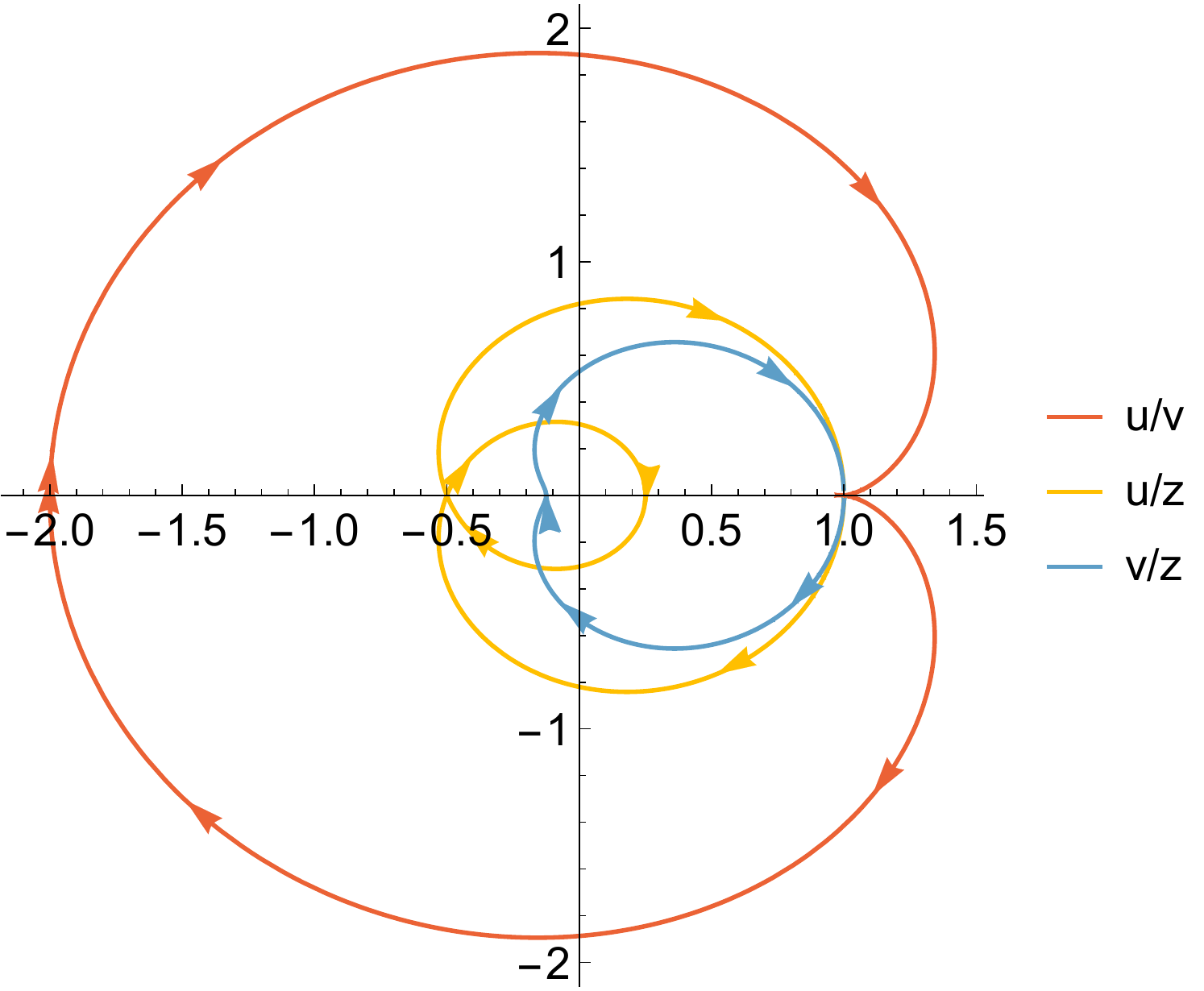}
	\caption{The trajectory followed by the various cross-ratios under analytic continuation to Lorentzian times. To make this figure we chose $\beta=2$, $\sigma_W=0.5$, $\sigma_Y=0.25$, and $\sigma_X=0$, as well as  $t_X=2t_Y$, with $t_W=0$. The anti-holomorphic cross-ratios do not cross any branch cuts. }
	\label{fig:cuts}
\end{figure}

To diagnose fine-grained chaos in this way, we use the exponential map to transform our vacuum blocks to thermal blocks for a CFT on the line: 
\begin{equation}
 	x_i\rightarrow e^{\frac{2\pi}{\beta} \tilde{x}_i}~,\quad\quad y_i\rightarrow e^{\frac{2\pi}{\beta} \tilde{y}_i}\quad\quad w_i\rightarrow e^{\frac{2\pi}{\beta} \tilde{w}_i}~.
 \end{equation} 
We will be interested in the following arrangement: 
\begin{align}
&\tilde{x}_i=t_X+\sigma_X-i\varepsilon_{X_i}~,& &\bar{\tilde{x}}_i=-t_X+\sigma_X+i\varepsilon_{X_i}~,\\
&\tilde{y}_i=t_Y+\sigma_Y-i\varepsilon_{Y_i}~,& &\bar{\tilde{y}}_i=-t_Y+\sigma_Y+i\varepsilon_{Y_i}~,\\
&\tilde{w}_i=t_W+\sigma_W-i\varepsilon_{W_i}~,& &\bar{\tilde{w}}_i=-t_W+\sigma_W+i\varepsilon_{W_i}~,
\end{align}
and to ensure the Lorentzian time ordering of interest, we will analytically continue from the Euclidean configuration starting at $t_X=t_Y=t_W=0$ with
\begin{equation}
 	\varepsilon_{X_1}>\varepsilon_{Y_1}>\varepsilon_{X_2}>\varepsilon_{W_1}>\varepsilon_{Y_2}>\varepsilon_{W_2}>0~.
 \end{equation} 
 In what follows, we will always take
 \begin{equation}
 	t_W<t_Y<t_X~,\qquad \text{and}\qquad \sigma_W>\sigma_Y>\sigma_X~.
 \end{equation}
 Under this analytic continuation to Lorentzian times, the relevant cross-ratios $z,u,v$ (and their ratios) cross the various branch cuts of the conformal blocks we have just computed \cite{Roberts:2014ifa}.  The trajectory followed by the relevant cross-ratios during the analytic continuation is depicted in figure \ref{fig:cuts}.  As is clear from \eqref{eq:Idef} (and less clear from \eqref{eq:combSol3}), the types of branch cuts crossed are of those of the logarithm, which extendes from $(-\infty,0]$ and the branch cut of the dilogarithm, which extends from $[1,\infty)$. These satisfy: 
 \begin{equation}
\begin{split}
  \text{for }z \in (-\infty,0]: &\qquad \lim_{\delta \rightarrow 0} \left[\,\log(z+i\delta) -\log(z-i\delta)\, \right] = 2\pi i    \,,\\
  \text{for } z \in [1,\infty): &\qquad  \lim_{\delta \rightarrow 0} \left[\,\text{Li}_2(z+i\delta) -\text{Li}_2(z-i\delta)\, \right] = 2\pi i \, \log z \,.
   \end{split}
\end{equation}

To obtain the six-point OTOC, we cross the various branch cuts of the blocks we computed, before taking the limit $z,u,v\rightarrow 1$. This is the Regge limit of these blocks. We can now present our result. 

\paragraph{The contribution considering only the global $T$ block:} Let us first give the result of the global block alone, $\Gstar$:
\begin{multline}
 	\text{OTOC}_{6pt}^{\text{star(global)}}\approx 1+\frac{96 i \pi h_Xh_Wh_Y}{c^2\,\varepsilon_{X_{12}}\varepsilon_{Y_{12}}\varepsilon_{W_{12}}}\left[\varepsilon_{X_{12}}^3\frac{\sinh^4\left(\frac{\pi(t_{YW}-\sigma_{WY}) }{\beta}\right)}{\sinh^2\left(\frac{\pi(t_{XY}-\sigma_{YX}) }{\beta}\right)\sinh^2\left(\frac{\pi(t_{XW}-\sigma_{WX}) }{\beta}\right)}\right.\\\left.+\varepsilon_{W_{12}}^3\frac{\sinh^4\left(\frac{\pi(t_{XY}-\sigma_{YX}) }{\beta}\right)}{\sinh^2\left(\frac{\pi(t_{YW}-\sigma_{WY}) }{\beta}\right)\sinh^2\left(\frac{\pi(t_{XW}-\sigma_{WX}) }{\beta}\right)}\right]
\end{multline}
where $t_{AB}=t_A-t_B$ and similarly for $\sigma_{AB}$ and $\varepsilon_{A_{12}}$. Note that the above block does not exhibit scrambling with the largest time difference $t_{XW}$. The contribution of this channel to the OTO correlator decays rather than grows, suggesting it is irrelevant for diagnosing chaos. 

\paragraph{The full OTOC:} We have learned in section \ref{sec:virasoro} that the Virasoro identity block has an additional contribution on top of the global block $\Gstar$, even at leading nontrivial order in large $c$. 
Let us therefore compute the leading contribution to the OTOC taking into account the Virasoro contributions in \eqref{eq:V6Result}:
\begin{equation}
\label{eq:virOTOC}
\boxed{ \;\;
\text{OTOC}_{6pt}^{\text{star(Virasoro)}}\approx -\frac{1152\,\beta^4 h_Xh_W h_Y}{c^2 \pi^2\varepsilon_{X_{12}}\varepsilon_{Y_{12}}^2\varepsilon_{W_{12}}}\sinh^2\left(\frac{\pi(t_{XY}-\sigma_{YX}) }{\beta}\right)\sinh^2\left(\frac{\pi(t_{YW}-\sigma_{WY}) }{\beta}\right) \;\;
}
\end{equation}
up to terms at ${\cal O}(\varepsilon_{ij}^{-3})$.
For large time separations $t_{XW} \gg t_{XY}  \sim t_{YW} \gg \frac{\beta}{2\pi}$, we notice that this term grows exponentially as $e^{\lambda_L t_{XW}}$, with 
\begin{equation}
	\lambda_L = \frac{2\pi}{\beta} \,.
\end{equation}
Furthermore, the star channel Virasoro OTOC becomes ${\cal O}(1)$ (and our approximations break down) when $t_{XW} \sim 2 \,t_\star$, i.e., the exponential growth lasts until {\it twice} the scrambling time $t_\star\equiv \frac{\beta}{2\pi}\log c$, as first noted in \cite{Haehl:2017pak,Haehl:2018izb}~. This was seen in those references as an indication that higher-point OTOCs contain novel information compared to the four-point OTOC: they are sensitive to more fine-grained aspects of quantum chaos  in the sense that their characteristic scrambling time grows (linearly) with the number of insertions. Also note that the butterfly velocity $v_B=1$ is the speed of light, as is typical for large-$c$ 2d CFT.

\section{Comparison with the global \texorpdfstring{$T$}{T} block in the comb channel}
\label{sec:comb}

So far we focused on the star channel blocks because these allow for the most natural six-point generalization of the familiar identity blocks. However, as mentioned in \S\ref{sec:blocks}, there also exists the comb topology (on the right of figure \ref{fig:starVScomb}), which also admits a universal contribution to the six-point function when the internal operator coincides with one of the external ones. In this section we study the comb channel block in some more detail and compare its properties with the star channel. For simplicity we shall only discuss the global comb block for which an explicit expression is available \cite{Rosenhaus:2018zqn}.

\paragraph{Definition of the global comb block:}
The universal comb channel six-point CPW is defined by projecting the fusion of $X$ and $W$ operators onto the identity representation using $|T|$ and inserting a projection $|Y|$ in the middle of the six-point function:
\begin{align}\label{eq:combDef}
  \Psicomb &\equiv\frac{\big{\langle} X(x_1)X(x_2) \,|T|\, Y(y_1) \,|Y|\, Y(y_2)\, |T|\, W(w_1) W(w_2) \big{\rangle}}{\langle VV\rangle \langle XX\rangle \langle WW \rangle }\nonumber\\
  &  =\frac{\Gamma(2h_Y)}{\pi\Gamma(1-2\bar{h}_Y)}  \left( \frac{3}{\pi c}\right)^2   \iiint d^2z_a\, d^2z_b\, d^2z_c\;  \frac{\big{\langle} X_1 X_2 {T}_a \big{\rangle}}{\langle X_1X_2\rangle}   \frac{\big{\langle}  \widetilde{T}_a Y_1 Y_b \big{\rangle}\big{\langle} \widetilde{Y}_b Y_2 T_c \big{\rangle}}{\langle Y_1Y_2\rangle}   \frac{\big{\langle}\widetilde{T}_c  W_1 W_2 \big{\rangle} }{\langle W_1W_2\rangle}
   \end{align}
where we abbreviate $X_1 \equiv X(x_1)$ etc.
This OPE channel is illustrated in the right panel of figure \ref{fig:starVScomb}. It was discussed more generally in \cite{Rosenhaus:2018zqn}, where comb channel blocks were determined explicitly in terms of generalizations of hypergeometric functions. See also \cite{Banerjee:2016qca,Alkalaev:2018nik,Kusuki:2019gjs,Anous:2019yku,Jensen:2019cmr} for physical applications and \cite{Alkalaev:2016rjl,Fortin:2019zkm,Parikh:2019ygo,Parikh:2019dvm,Jepsen:2019svc,Alkalaev:2020kxz} for discussions in the context of Witten diagrams and AdS/CFT.

\paragraph{Casimir equations:}
The comb channel block $\Gcomb(z,u,v)$ in the right of figure \ref{fig:starVScomb} has the $X$ and $W$ operators fusing into stress tensors, which results in $\Gcomb(z,u,v)$ also satisfying \eqref{eq:6ptCasimir1} and \eqref{eq:6ptCasimi3}. However, instead of \eqref{eq:6ptCasimi2} it must satisfy:
\begin{align}\label{eq:6ptCasimirComb}
  {\cal C}(x_1,x_2,y_1) \left[ \langle XX \rangle \langle YY \rangle \langle WW\rangle  \, \Psicomb  \right] &= h_Y(h_Y-1) \, \langle XX \rangle \langle YY \rangle \langle WW\rangle \, \Psicomb
\end{align}
In terms of cross-ratios, we have \eqref{eq:cas2}, \eqref{eq:cas3} and in addition the following equation:
\begin{multline}\label{eq:cas4}
  \bigg\lbrace v^2(1-v)\partial_v^2+u^2(1-u)\partial_u^2+uv[(1-u)+(1-v)]\partial_u\partial_v+(1-z)\left[v^2\partial_v+u^2\partial_u\right]\partial_z\\-v(v-2h_Y)\partial_v-u(u-2h_Y)\partial_u\bigg\rbrace \, \Gcomb(z,u,v) =  0
\end{multline}

\paragraph{Explicit form of the comb block:}
The comb channel expression \eqref{eq:combDef} was worked out in \cite{Rosenhaus:2018zqn} (see also \cite{saran1954hypergeometric,saran1955solutions,saran1955transformations,pandey1963certain} for earlier papers where the relevant hypergeometric functions make an appearance). Writing their result in terms of our cross-ratios \eqref{eq:crossDef}, we obtain:
\begin{equation} 
\label{eq:combSol1}
\begin{split}
  \Gcomb(z,u,v)
    & =\frac{2\, h_Xh_W h_Y^2}{c^2}\, \frac{(u-v)^2(1-z)^2}{v^2 z^2} \;  F_K  \left[ \begin{aligned} 2,\,2,\,2,\,2 \\ 4, 2h_Y , 4 \end{aligned} \, \bigg{|} 1-\frac{1}{z} ,\; \frac{u}{z} ,\; 1- \frac{u}{v} \right] 
\end{split}
\end{equation}
where the hypergeometric function of three variables is defined as: 
{\small
\begin{equation}\label{eq:FKdef}
\begin{split}
 F_K \left[ \begin{aligned} a_1,\,b_1,\,b_2,\,a_2 \\ c_1, c_2 , c_3\quad \end{aligned} \, \bigg{|} \, \chi_1, \chi_2, \chi_3 \right] 
  &\equiv \sum_{n_1,n_2,n_3 = 0}^\infty \frac{ (a_1)_{n_1} (b_1)_{n_1+n_2} (b_2)_{n_2+n_3} (a_2)_{n_3} }{(c_1)_{n_1} (c_2)_{n_2} (c_3)_{n_3}} \, \frac{\chi_1^{n_1}}{n_1!} \frac{\chi_2^{n_2}}{n_2!} \frac{\chi_3^{n_3}}{n_3!} \,.
\end{split}
\end{equation}
}\normalsize
For the values of $(a_i,b_i,c_i)$ appearing in \eqref{eq:combSol1}, the hypergeometric function can be further simplified. We provide some of the steps in Appendix \ref{app:hypergeometric}. After some simplifications, the final result can be written in terms of single-variable hypergeometric functions: 
\begin{equation} 
\label{eq:combSol2}
\boxed{
    \;\;\Gcomb(z,u,v) = \frac{36\, h_Xh_Y h_W}{c^2} \left[ {\cal J}(z,u,v)  + {\cal J}\left( \frac{1}{z}, \frac{u}{z}, \frac{v}{z} \right)  \right] \;\;
}
\end{equation}
where we defined the function
{\small
\begin{equation} 
\label{eq:combSol3}
\begin{split}
&{\cal J}(z,u,v)  \equiv \frac{u}{2} \left( \frac{u+v}{u} - \frac{2v}{u-v} \, \log  \frac{u}{v} \right) \left( \frac{1+z}{z} +\frac{2}{1-z} \,\log z \right)+  h_Y \left( 2 - \frac{u+v}{u-v}  \log \frac{u}{v} \right) \left( 2 + \frac{1+z}{1-z} \log z \right)  \\
&  \qquad\qquad\;\; +  \frac{1}{2h_Y+1} \bigg\{ \big[F_1(u) +F_1(v) \big] -  \frac{4uv}{(u-v)(1-z)} \, \big[ F_2(u) -F_2(v)  \big]  +  \frac{(u+v)(1+z)}{2(u-v)(1-z)} \, \big[F_3(u) -F_3(v) \big] \bigg\}
\end{split}
\end{equation}
}\normalsize
where for ease of notation, we abbreviate
\begin{align}
F_1(\chi)&\equiv \chi^2\, {}_2F_1(1,1;\, 2h_Y+2;\, \chi)\\
F_2(\chi)&\equiv \chi \;  {}_3F_2(1,1,1;\,2,2h_Y+2;\,\chi)\\
F_3(\chi)&\equiv \chi^2\; {}_3F_2(1,1,2;\,3,2h_Y+2;\,\chi)
\end{align}
It can be checked that, for $h_Y\in \mathbb{Z}$, these functions reduce to polynomials multiplying logs and dilogs. Therefore, similar to the star channel block, this result displays a rich structure of branch cuts.
Note also that the first line of \eqref{eq:combSol3} can be recognized as being closely related to the disconnected product of two four-point blocks. It is straightforward to check that the above expression satisfies the required Casimir equations.

\begin{figure}
	\centering     
	\includegraphics[width=.25\textwidth]{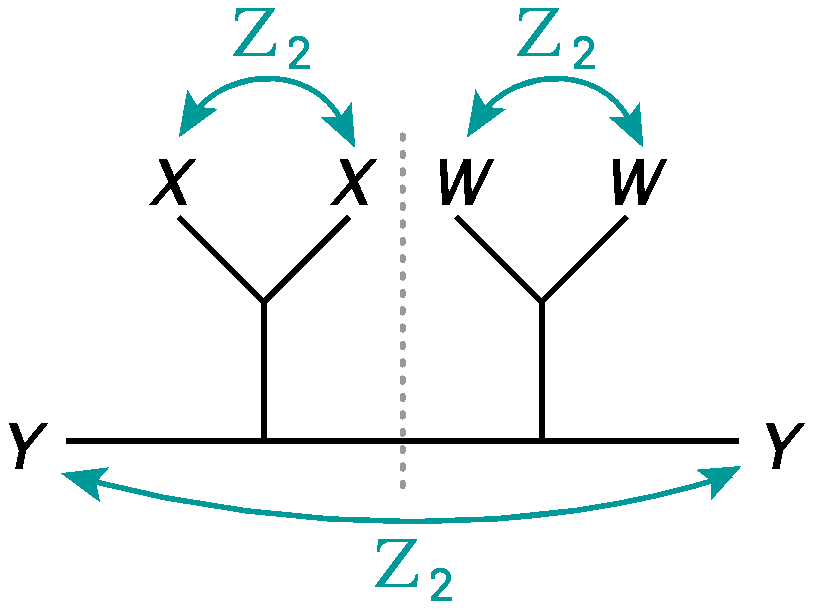}
	\caption{Permutation symmetries of the comb channel block.}
	\label{fig:symms2}
\end{figure}

\paragraph{Symmetry properties:}
The symmetries of the comb channel block are slightly different to the star channel. They can be summarized by the group $\big(\mathbb{Z}_2^{(X)} \times \mathbb{Z}_2^{(W)} \big)  \rtimes \mathbb{Z}_2^{\text{(refl.)}}$, where the first two factors are the same as for the star channel, and the ``reflection'' $\mathbb{Z}_2^{\text{(refl.)}}$ acts by $(X_1,X_2,Y_1) \leftrightarrow (W_1,W_2,Y_2)$.\footnote{ At the level of cross-ratios, $\mathbb{Z}_2^{\text{(refl.)}}$ acts as: $(z,u,v) \mapsto \left(\frac{u}{v},u,\frac{u}{z}\right)$.} We illustrate these symmetries schematically in figure \ref{fig:symms2}.

\paragraph{Physical significance:} Let us now comment on the physical significance of this global block. Recall that in the previous section we found that growing contribution to the OTO correlator at six points in the star channel came from the contribution $\Gext$ to the block and \emph{not} $\Gstar$.  
We thought it prudent to compare the contribution of $\Gext$ to the comb block. Namely, we assume there exists a comb channel Virasoro identity block $\Vcomb$ whose leading large-$c$ expansion is
\begin{equation}
	\Vcomb\approx 1+ \Gcomb+\dots
\end{equation}
We will ignore the possibility that there are additional terms at the same order, though it would be interesting to explore this possibility further. We leave the exploration of these additional terms to future work, as it would require developing novel techniques, beyond the scope of our current setup, for computing them. Under this assumption we can compute the six-point OTOC and find:
\begin{equation}
\label{eq:combOTOC}
\text{OTOC}_{6pt}^{\text{comb}}\approx -\frac{576\,\beta^4 h_Xh_W h_Y(2h_Y+1)}{c^2 \pi^2\varepsilon_{X_{12}}\varepsilon_{Y_{12}}^2\varepsilon_{W_{12}}}\sinh^2\left(\frac{\pi(t_{XY}-\sigma_{YX}) }{\beta}\right)\sinh^2\left(\frac{\pi(t_{YW}-\sigma_{WY}) }{\beta}\right)
\end{equation}
which displays the same late-time dependence as in the case of the star channel Virasoro block \eqref{eq:virOTOC}.

\section{Conclusions and outlook}
\label{sec:conclusions}

The ideas presented here combine two topics of recent interest. 
The first topic is the conformal block decomposition of correlation functions in large-$c$ CFTs. Recent developments have shown how topologically distinct Virasoro conformal block expansions can exchange dominance in describing a certain correlation function \cite{Hartman:2013mia,Asplund:2014coa,Anous:2016kss,Anous:2017tza}. For  the case of six-point functions, the topological distinction is borne out even in the global blocks, which are distinguished by a choice of OPE channel.

The second topic of the paper was to provide some further developments in the shadow operator formalism for stress tensor exchanges and its nonlinear generalization. The close connection with recent studies of theories of reparametrization modes in CFTs offers a powerful framework for computing observables such as the star channel six-point block, and higher order interactions.

 The optimal choice of block decomposition, measured by its relevance to a physical process, can only be determined by comparison. In this paper we found explicit expressions for global six-point blocks with internal stress tensor exchanges in the comb and in the star channel, as well as additional terms that contribute to the star channel Virasoro block at the same order as the global block. While these have interesting commonalities (such as multiple branch cuts of higher transcendentality), we found that only the non-global terms in the star channel, as well as the global comb channel give the leading contribution to scrambling. Interestingly, the star channel global block, which captures a nonlinear gravitational three-point interaction in the holographic dual, does not seem relevant for the connected six-point OTOC \cite{Haehl:2017pak,Haehl:2018izb}. Nevertheless, it may give insights into subtle gravitational physics, which we hope to explore further.

 It is interesting that the star channel Virasoro identity block in various kinematic regimes receives leading contributions other than the global block, and these correspond to multiple linearized graviton exchanges. The fact that these are not only present but in fact crucial for quantum chaos was anticipated in \cite{Haehl:2017pak,Haehl:2018izb}. If we interpret these results in terms of an effective field theory of chaos described by reparametrization mode exchanges, the lesson is similar: self-interactions are less relevant for quantum chaos than multiple connected linearized exchanges \cite{Fitzpatrick:2015qma,Kulaxizi:2017ixa,Kulaxizi:2018dxo}. Furthermore, in the comb channel, we observe that the global block is sufficient  for diagnosing chaos, but nevertheless there could be additional contributions appearing at the same order which we have not yet computed, in direct analogy with the piece $\Gext$ that appears in the star channel. If such contributions exist, we expect them to not affect the time dependence of the chaotic growth of the OTO configuration. It would be interesting to understand the general lesson behind these observations and see if the methods developed herein can be useful to the understanding of eikonlization, even in higher dimensions \cite{Kulaxizi:2019tkd,Fitzpatrick:2019efk}.  Perhaps we need to look beyond chaos to diagnose nonlinearities in gravity.

In the future, we would also like to explore how to generalize these results to full Virasoro six-point blocks. One block of particular interest is the semiclassical ``HHLLLL block" with two heavy operators ($h_Y \sim\mathcal{O}(c)$) and four light probes ($h_X \sim h_W \sim {\cal O}(c^0)$). This was computed in a particular regime in \cite{Anous:2019yku} (see also \cite{Jensen:2019cmr}) in the semiclassical limit where all operator dimensions scale linearly with $c$, with the light operators having small $h_{\{X,Y\}}/c$. Interestingly, the large-$c$ expansion of the result of \cite{Anous:2019yku} only gives a piece of the global star block computed here, without the dilogarithms. It would be interesting to understand how to obtain the full star block as a limit of the monodromy method for heavy-light correlators. This would allow us to more finely probe a black hole microstate with four operators, in the geometric optics limit, sensitive to the nonlinear gravitational self-interaction. 

Based on insights for the four-point HHLL block \cite{Fitzpatrick:2015zha}, there is a natural guess for what the HHLLLL Virasoro six-point identity block would be: we conjecture that it is given by the leading block $\Gstar+\Gext$ presented in \S\ref{subsec:light} evaluated in the coordinates $w(x) = x^\alpha$ where $x=z,u,v$ and $\alpha = \sqrt{1-24 h_Y/c}$ (see eq.\ (1.3) of \cite{Fitzpatrick:2015zha}). While this seems natural, it would be interesting to verify it in detail.

\acknowledgments
The authors are indebted to Eric Perlmutter for initial collaboration and many enlightening discussions. We also thank Pawel Caputa, Eliot Hijano, Wyatt Reeves, Moshe Rozali, and Alexandre Streicher for stimulating conversations and comments. TA thanks the koffietijd groep for emotional support. TA is supported by the Delta ITP consortium, a program of the Netherlands Organisation for Scientific Research (NWO) that is funded by the Dutch Ministry of Education, Culture and Science (OCW). TA also acknowledges support from the KITP at UC Santa Barbara and the National Science Foundation under Grant No. NSF PHY-1748958. FH acknowledges support by the US Department of Energy under grant DE-SC0009988.

\appendix

\section{Six-point Virasoro block for `hefty' operators}
\label{app:virasoro}

In this appendix we give the derivation of the result that the star channel Virasoro identity block is trivial at leading order, assuming operator dimensions scale as $c^{1/2}$, i.e., the fact that \eqref{eq:VirRes} has no nontrivial contribution at order ${\cal O}(c^0)$.

Translating the definition \eqref{eq:V6def} into diagrams, at leading order we need to sum all diagrams of the form presented in figure \ref{fig:kl}.
We begin by computing the simplest ingredient in \eqref{eq:V6def}, i.e., the one-point functions. For ease of notation we will henceforth drop the explicit spacetime arguments of the bilocals. We find:
\begin{equation}
\begin{split} 
 \big\langle {\cal B}_{\epsilon,h_X} \big\rangle_{{\cal O}(c^0)} &= \left\langle \exp {\cal B}^{(1)}_{\epsilon,h_X} \right\rangle 
   = \sum_{\ell_1 \geq 0}   \;\;\begin{gathered} \includegraphics[width=.07\textwidth]{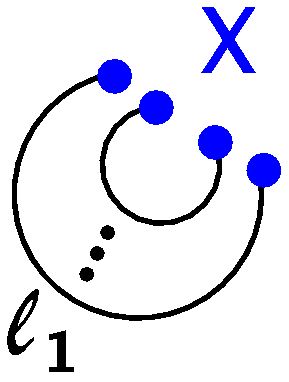} \end{gathered} \\
   &= \sum_{\ell_1 \geq 0} \frac{1}{(2\ell_1)!} \, (2\ell_1-1)!! \, \left\langle\left( {\cal B}^{(1)}_{\epsilon,h_X}\right)^2 \right\rangle^{\ell_1}
   = \exp \left[ \frac{1}{2} \left\langle\left( {\cal B}^{(1)}_{\epsilon,h_X}\right)^2 \right\rangle\right]
\end{split}
\end{equation}
where all equalities hold up to subleading terms of order ${\cal O}(c^{-1/2})$.
The factor $(2\ell_1-1)!!$ is a symmetry factor counting the number of ways to contract $2\ell_1$ copies of ${\cal B}^{(1)}_{\epsilon,h_X}$ in pairs.

Next, we turn to the four-point functions in the denominator of \eqref{eq:V6def} (this part of the calculation is similar to one presented in \cite{Cotler:2018zff}). We find:
{\small
\begin{equation}
\label{eq:B1B1exponential}
\begin{split} 
 &\big\langle {\cal B}_{\epsilon,h_X} \,{\cal B}_{\epsilon,h_Y} \big\rangle_{{\cal O}(c^0)}
   = \left\langle \exp \left( {\cal B}^{(1)}_{\epsilon,h_X} \right) \; \exp \left( {\cal B}^{(1)}_{\epsilon,h_Y} \right)\right\rangle
   = \sum_{\ell_1 \geq 0} \sum_{\ell_2 \geq 0} \sum_{k_1 \geq 0} \;\;\begin{gathered}\vspace{-.1cm} \includegraphics[width=.1\textwidth]{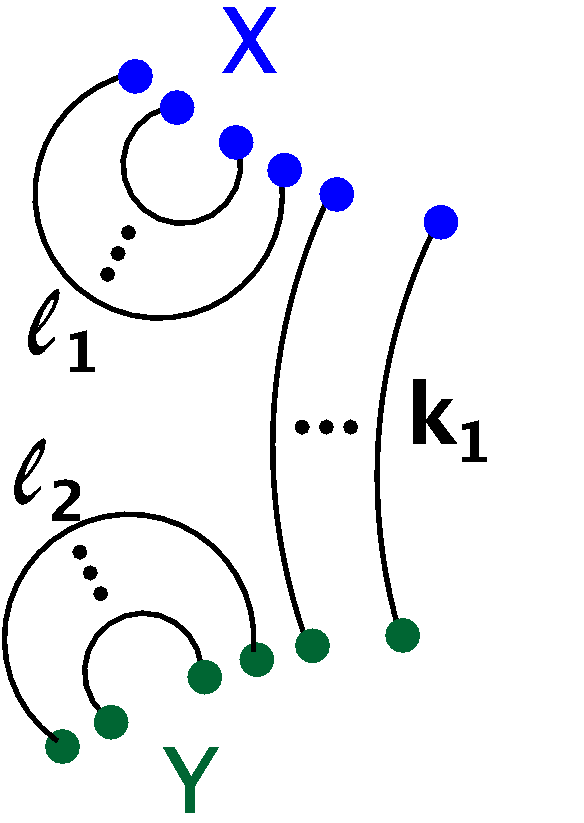} \end{gathered} \\
   &\quad= \sum_{\ell_1 \geq 0} \sum_{\ell_2 \geq 0} \sum_{k_1 \geq 0}   \frac{ {2\ell_1+k_1 \choose k_1 } {2\ell_2+k_1 \choose k_1 }  (2\ell_1-1)!!  \, (2\ell_2-1)!! \, k_1!}{(2\ell_1+k_1)! (2\ell_2+k_1)!} \;  \left\langle {\cal B}^{(1)}_{\epsilon,h_X} \, {\cal B}^{(1)}_{\epsilon,h_Y} \right\rangle^{k_1}\, \left\langle\left( {\cal B}^{(1)}_{\epsilon,h_X}\right)^2 \right\rangle^{\ell_1} \,  \left\langle\left( {\cal B}^{(1)}_{\epsilon,h_Y}\right)^2 \right\rangle^{\ell_2}  \\
   &\quad= \exp \left[   \left\langle {\cal B}^{(1)}_{\epsilon,h_X} \, {\cal B}^{(1)}_{\epsilon,h_Y} \right\rangle + \frac{1}{2} \,  \left\langle\left( {\cal B}^{(1)}_{\epsilon,h_X}\right)^2 \right\rangle + \frac{1}{2} \,  \left\langle\left( {\cal B}^{(1)}_{\epsilon,h_Y}\right)^2 \right\rangle  \right]
\end{split}
\end{equation}
}\normalsize
where we again included appropriate symmetry factors, this time also doing the binomial counting of possible numbers of ways to choose $k_1$ out of $(2\ell_1+k_1)$ copies of ${\cal B}^{(1)}_{\epsilon,h_X}$ (and similarly for ${\cal B}^{(1)}_{\epsilon,h_Y}$).

Finally, the three-point function in the numerator of \eqref{eq:V6def} gives the following: 
{\small
\begin{equation}
\begin{split} 
 &\big\langle {\cal B}_{\epsilon,h_X} \,{\cal B}_{\epsilon,h_Y}\,{\cal B}_{\epsilon,h_W} \big\rangle_{{\cal O}(c^0)} 
   = \left\langle \exp \left( {\cal B}^{(1)}_{\epsilon,h_X} \right) \; \exp \left( {\cal B}^{(1)}_{\epsilon,h_Y} \right)\; \exp \left( {\cal B}^{(1)}_{\epsilon,h_W} \right)\right\rangle
   = \sum_{\substack{\ell_1,\ell_2,\ell_3\\ k_1,k_2,k_3}}\;\;\begin{gathered} \includegraphics[width=.17\textwidth]{SixPtVirNew.png} \end{gathered} \\
   &\quad= \sum_{\substack{\ell_1,\ell_2,\ell_3\\ k_1,k_2,k_3}}  \frac{ {2\ell_1+k_1+k_3 \choose k_3 } {2\ell_1+k_1 \choose k_1 } {2\ell_2+k_2+k_1 \choose k_1 } {2\ell_2+k_2 \choose k_2 } {2\ell_3+k_3+k_2 \choose k_2 } {2\ell_3+k_3 \choose k_3 } (2\ell_1-1)!! \, (2\ell_2-1)!! \, (2\ell_3-1)!! \, k_1! \, k_2! \, k_3! }{(2\ell_1+k_1+k_2)!\,(2\ell_2+k_2+k_3)!\,(2\ell_3+k_3+k_1)!}  \\
   &\qquad\qquad\quad\; \times 
       \left\langle {\cal B}^{(1)}_{\epsilon,h_X} \, {\cal B}^{(1)}_{\epsilon,h_Y} \right\rangle^{k_1}\!
       \left\langle {\cal B}^{(1)}_{\epsilon,h_Y} \, {\cal B}^{(1)}_{\epsilon,h_W} \right\rangle^{k_2}\!
       \left\langle {\cal B}^{(1)}_{\epsilon,h_W} \, {\cal B}^{(1)}_{\epsilon,h_X} \right\rangle^{k_3}\!
        \Big\langle\left( {\cal B}^{(1)}_{\epsilon,h_X}\right)^2 \Big\rangle^{\ell_1} \!
         \Big\langle\left( {\cal B}^{(1)}_{\epsilon,h_Y}\right)^2 \Big\rangle^{\ell_2}  \!
         \Big\langle\left( {\cal B}^{(1)}_{\epsilon,h_W}\right)^2 \Big\rangle^{\ell_3}\\
   &\quad= \exp \left[  
    \left\langle {\cal B}^{(1)}_{\epsilon,h_X} \, {\cal B}^{(1)}_{\epsilon,h_Y} \right\rangle 
    +  \left\langle {\cal B}^{(1)}_{\epsilon,h_Y} \, {\cal B}^{(1)}_{\epsilon,h_W} \right\rangle  
    +  \left\langle {\cal B}^{(1)}_{\epsilon,h_W} \, {\cal B}^{(1)}_{\epsilon,h_X} \right\rangle \right.\\
    &\qquad\qquad \left.
     + \frac{1}{2} \,  \Big\langle\left( {\cal B}^{(1)}_{\epsilon,h_X}\right)^2 \Big\rangle + \frac{1}{2} \,  \Big\langle\left( {\cal B}^{(1)}_{\epsilon,h_Y}\right)^2 \Big\rangle+ \frac{1}{2} \,  \Big\langle\left( {\cal B}^{(1)}_{\epsilon,h_W}\right)^2 \Big\rangle  \right] 
\end{split}
\end{equation}
}\normalsize

Putting these pieces together we find that the Virasoro identity block with these scalings and at this order is just 1. All nontrivial dependence is cancelled by the normalization in \eqref{eq:V6def}. This is a nontrivial confirmation of our claim that all disconnected diagrams are cancelled.

\section{Simplification of the comb channel block}
\label{app:hypergeometric}

In this appendix we provide some details on simplifying the hypergeometric function \eqref{eq:FKdef} to obtain the comb channel block \eqref{eq:combSol2}. The hypergeometric function can be written as a sum over lower order hypergeometric functions:
{\small
\begin{equation}\label{eq:FKdef2}
\begin{split}
 F_K \left[ \begin{aligned} a_1,\,b_1,\,b_2,\,a_2 \\ c_1, c_2 , c_3\quad \end{aligned} \, \bigg{|} \, \chi_1, \chi_2, \chi_3 \right] 
   &= \sum_{n=0}^\infty \frac{ (b_1)_n(b_2)_n }{(c_2)_n} \,  \frac{\chi_2^{n}}{n!} \; {}_2F_1(b_1+n, \, a_1,\,c_1 ; \chi_1)\; {}_2F_1(b_2+n, \, a_2,\,c_3 ; \chi_3) \,.
\end{split}
\end{equation}
}\normalsize
For the values of $(a_i,b_i,c_i)$ appearing in \eqref{eq:combSol1}, this can be further simplified. Note that:
{\small
\begin{equation}
\label{eq:tempSum}
\begin{split}
 & F_K  \left[ \begin{aligned} 2,\,2,\,2,\,2 \\ 4, 2h , 4 \end{aligned} \, \bigg{|}  \chi_1,\chi_2,\chi_3 \right] \\
 &\quad = \frac{36}{\chi_1^3 \chi_3^3} \sum_{n=0}^\infty \frac{ n! }{(2h)_n} \,  \frac{\chi_2^{n}}{n^2(n-1)^2} \left( (2-\chi_1 +n\chi_1) - \frac{2-\chi_1-n\chi_1}{(1-\chi_1)^n} \right)\left( (2-\chi_3 +n\chi_3) - \frac{2-\chi_1-n\chi_3}{(1-\chi_3)^n} \right) 
\end{split}
\end{equation}
}\normalsize
The terms $n=0,1$ can be evaluated directly and give finite logarithmic terms. We can consider each term in the remaining sums separately and write it in terms of hypergeometric functions. To this end, we write the two $n$-dependent brackets as follows:
\small{
\begin{equation}
\label{eq:nIdent}
\begin{split}
&\left( (2-\chi_1 +n\chi_1) - \frac{2-\chi_1-n\chi_1}{(1-\chi_1)^n} \right)\left( (2-\chi_3 +n\chi_3) - \frac{2-\chi_1-n\chi_3}{(1-\chi_3)^n} \right)  \\
&\;\; = (1+\eta_1)(1+\eta_3)\left(1-\eta_1^{-n}\right) \left(1-\eta_3^{-n}\right) \\
&\quad + n   \left[ 2\left( 1+\eta_1^{-n}\right)\left( 1-\eta_3^{-n}\right) + 2\left( 1-\eta_1^{-n}\right)\left( 1+\eta_3^{-n}\right)
   - (1-\eta_1)(1-\eta_3) \left( 1 - \eta_1^{-n} - \eta_3^{-n} - 3\, (\eta_1\eta_3)^{-n} \right) \right] \\
 &\quad + n(n-1) \, (1-\eta_1)(1-\eta_3)  \left(1+\eta_1^{-n}\right)\left(1+\eta_3\right)^{-n} \,.
\end{split}
\end{equation}
}\normalsize
where $\eta_i \equiv 1-\chi_i$.
Then, note the following identities:
\begin{equation}
\label{eq:sums}
\begin{split}
 \sum_{n=2}^\infty \frac{ n! }{(2h)_n} \,  \frac{\chi^{n}}{n(n-1)} & = \frac{1}{2h(2h+1)} \; \chi^2 \; {}_2F_1(1,1;\, 2h+2; \, \chi) \\
      \sum_{n=2}^\infty \frac{ n! }{(2h)_n} \,  \frac{\chi^{n}}{n(n-1)^2}  &= \frac{1}{2h(2h+1)} \; \chi^2 \; {}_3F_2(1,1,1;\, 2,2h+2; \, \chi) \\
  \sum_{n=2}^\infty \frac{ n! }{(2h)_n} \,  \frac{\chi^{n}}{n^2(n-1)^2}  &= \frac{1}{4h(2h+1)} \; \chi^2 \; {}_3F_2(1,1,1;\, 3,2h+2; \, \chi)  
\end{split}
\end{equation}
Using \eqref{eq:nIdent} in \eqref{eq:tempSum}, the identities \eqref{eq:sums} allow us to perform all the sums. We further use the contiguous relation ${}_3F_2(1,1,1;\,3,a;\,\chi) = 2\;{}_3F_2(1,1,1;\,2,a;\,\chi) - {}_3F_2(1,1,2;\,3,a;\,\chi)$. The result is \eqref{eq:combSol2}. The first line of \eqref{eq:combSol2} corresponds to the terms $n=0,1$ of the sum \eqref{eq:tempSum}.

\paragraph{Properties of the comb block hypergeometric functions:}
Note that the generalized hypergeometric functions have the following structure for integer values of $h$:
\begin{equation}
\begin{split}
   {}_2F_1(1,1;\, 2h+2; \, \chi)  &=  - (2h+1) \frac{(1-\chi)^{2h}}{\chi^{2h+1}} \, \log(1-\chi) + \ldots \\
     {}_3F_2(1,1,1;\, 2,2h+2; \, \chi)  & = (2h+1) \; \frac{1}{\chi} \, \text{Li}_2(\chi) + (\ldots) \, \log(1-\chi) + \ldots  \\
   {}_3F_2(1,1,1;\, 3,2h+2; \, \chi)&  = 2 \left( 2h+\chi\right)(2h+1)\,\frac{1}{\chi^2}\; \text{Li}_2(\chi)  + (\ldots) \, \log(1-\chi) + \ldots
\end{split}
\end{equation} 
where dots denote rational functions of $\chi$.

For the OTOC calculations, we need the discontinuity of the hypergeometric functions across their branch cut:
\begin{equation}
\label{eq:branch3F2}
\begin{split}
   &\lim_{\delta \rightarrow 0} \left[\, {}_3F_2(1,1,1;\,c_1,c_2;\,z+i\delta) - {}_3F_2(1,1,1;\,c_1,c_2;\,z-i\delta)\, \right]\\
   &\quad = 2\pi i \, \frac{\Gamma(c_1)\Gamma(c_2)}{\Gamma(c_1+c_2-2)} \, z^{2-c_1-c_2} (z-1)^{c_1+c_2-3} \; {}_2F_1(c_1-1,\, c_2-1,\, c_1+c_2-2; \, 1- z^{-1} ) \,.
   \end{split}
\end{equation}
for $z\in (1,\infty)$.
Relevant for us are values $c_1 \in \{1,2,3\}$.

\bibliographystyle{utphys}
\bibliography{EE}

\end{document}